\documentclass[aps,prb,preprint]{revtex4}
\usepackage[dvips]{graphicx}
\usepackage[dvips]{color}
\usepackage{amsmath}
\usepackage[below]{placeins}

\begin{document}

\pagestyle{plain}
\pagenumbering{arabic}

\title{Computer Simulation Study of the Phase Behavior and Structural Relaxation in 
a Gel-Former Modeled by Three Body Interactions}
\author{Shibu Saw$^{1}$, Niels  L. Ellegaard$^{1}$, Walter Kob$^{2}$, Srikanth Sastry$^{1}$}
\affiliation{$^{1}$ Theoretical Sciences Unit, Jawaharlal Nehru Centre for Advanced Scientific Research, Jakkur Campus, Bangalore 560 064, India.\\
$^{2}$ Laboratoire des Collo\"ides, Verres et Nanomat\'eriaux, UMR5587 CNRS, Universit\'e Montpellier 2, 34095 Montpellier, France}

\date{\today}
\widetext

\begin{abstract}
\begin{center}
\vspace{3mm}

\parbox{6in}{We report a computer simulation study of a model gel-former 
obtained by modifying the three-body interactions of the Stillinger-Weber
potential for silicon. This modification reduces the average coordination
number and consequently shifts the liquid-gas phase coexistence curve
to low densities, thus facilitating the formation of gels without phase
separation. At low temperatures and densities, the structure of the system
is characterized by the presence of long linear chains interconnected by a
small number of three coordinated junctions at random locations. At small
wave-vectors the static structure factor shows a non-monotonic dependence
on temperature, a behavior which is due to the competition between the
percolation transition of the particles and the stiffening of the formed
chains. We compare in detail the relaxation dynamics of the system as
obtained from molecular dynamics with the one obtained from Monte Carlo
dynamics. We find that the bond correlation function displays stretched
exponential behavior at moderately low temperatures and densities, but
exponential relaxation at low temperatures. The bond lifetime shows
an Arrhenius behavior, independent of the microscopic dynamics. For the
molecular dynamics at low temperatures, the mean squared displacement and
the (coherent and incoherent) intermediate scattering function display
at intermediate times a dynamics with ballistic character and we show
that this leads to compressed exponential relaxation. For the Monte
Carlo dynamics we find always an exponential or stretched exponential
relaxation. Thus we conclude that the compressed exponential relaxation
observed in experiments is due to the out-of-equilibrium dynamics.}

\end{center}  
 
\end{abstract}
\maketitle

\section{Introduction} 
Gels are ubiquitous in daily life including food, cosmetics and
medicines. They are low density disordered networks of interacting
molecules which can sustain weak stress.  In this sense they behave
like solids\cite{Manley-prl}. In spite of their low density,
fluids that form gels exhibit slow relaxation dynamics at low
temperatures, similar to glasses.  Based on the life time of bonds
between constituent units, gels can be classified as chemical or
physical gels\cite{Zacca,Cip04a,delgado,hurtado,hurtado2,Blaak,Russo}.
While chemical gels are formed by the formation of strong covalent
bonds in a disordered structure, physical gels (such as gelatin,
colloidal gels, etc.) arise due to relatively weak interactions.
The mechanisms for the formation of such physical gels, i.e.~gelation,
are still not fully understood. Gelation can arise in phase separated
solutions\cite{Manley-prl,soga,Lu-nature} due to the intersection
of the glass transition and spinodal lines \cite{sastry-prl-2000},
by the formation of a system spanning network in homogeneous
suspensions\cite{KobEuro,Zaccarelli-prl}, or due to aggregation of
small clusters by the process of diffusion limited cluster aggregation
\cite{bibette,suarez}.

Gels exhibit unusual dynamical properties. Their relaxation process is
often stretched exponential\cite{suarez}, a compressed
exponential\cite{Cipelletti,Bandyopadhyay}, or logarithmic\cite{Kroy}.
Experiments on non-equilibrium colloidal gels show compressed
exponential relaxation behavior of the dynamic structure factor with
an exponent of $\sim 1.5$ \cite{Cipelletti}. How these features are
related to the equilibrium properties of the gel-former and to the
details of the interactions between the particles is presently not
known. Unfortunately there exist so far only relatively few studies of
suspensions in equilibrium which approach gelation while remaining
homogeneous\cite{KobEuro,KobPRL,fierro,Russo,Zaccarelli-prl,Bianchi}.
In order to obtain such gels it is necessary for the system not to
enter the liquid-gas (LG) coexistence region on its approach to the
gelation line. This can be accomplished by shifting the liquid-gas
coexistence region to lower densities and temperatures, thereby
opening up a low temperature, low density region of the phase diagram
where the solution remains homogeneous but is characterized by the
presence of bonds which are strong relative to thermal
fluctuations. One way to effect such a shift is to reduce the maximum
coordination number which a particle can have\cite{fsrev}. A shifting
of the LG coexistence region in this manner has been realized for
patchy colloidal particle suspensions (where colloidal particles have
a small number of attractive patches on its
surface)\cite{KobEuro,Bianchi}, in maximum valency lattice gas
models\cite{Sastry-JSM}, hard sphere models with short range square
well attractive interactions\cite{Zaccarelli-prl}, and models with
dipolar interactions \cite{Blaak}. Although this scenario has not been
routinely realized experimentally, a recent study of colloidal clay
(laponite) presents an experimental realization of the above scenario \cite{lapo}.

The reduction in maximum coordination number can also be obtained
using three body interaction potentials\cite{KobEuro}. Three body interactions are
expressible in terms of angles formed by triplets of particles, and
can be chosen to energetically stabilize open network structures in
which the particles have very small (tunable) connectivity. One recent
example of this approach has been presented in \cite{shibuPRL}, wherein
the Stillinger-Weber (SW)\cite{SW} model potential for silicon has been
modified to generate structures with low coordination.

In the present paper, we study the static and dynamic properties of
gel-forming fluids obtained by the above-mentioned modification of the SW
interaction potential. Note that all the presented results are for the
system {\it in equilibrium}, i.e. they are not affected by aging phenomena.
The remaining of the paper is organized as follows:
In Sec. II, we describe the SW model potential of silicon and its
modification for homogeneous gelation, and provide the computational
details relevant to this work. In Sec. III, we discuss the static
properties of the system and the dynamic properties in Sec. IV. Finally,
Sec. V contains a summary and conclusions of our work.

\section{Modification of Stillinger-Weber  potential and computational details}

Here we describe the SW\cite{SW} potential which we modify in this
work. The SW potential was originally proposed with parameters chosen
to provide a reasonable description of the experimentally observed
thermodynamic and structural properties of crystalline and liquid
silicon. Its functional form is given by two-body and three-body
interaction terms and is written as

\begin{eqnarray}
u_{SW} = \sum_{i<j}v_2(r_{ij}/\sigma) + 
\sum_{i<j<k} v_3({\bf r}_i/\sigma,{\bf r}_j/\sigma,{\bf r}_k/\sigma), 
\label{eq1}
\end{eqnarray}

\noindent
where $\sigma$ is the diameter of the particles, ${\bf r}_i$ is the position
of particle $i$, and $r_{ij}$ is the distance between particles $i$
and $j$.  The two-body potential is  short-ranged and has the form

\begin{eqnarray}
v_2(r) &=&
\left \{
\begin{array}{ll}
A \epsilon (B r^{-4}-1)\exp{\left(\frac{1}{r-a}\right)} & r <a  \\
0                                                     & \geq a  
\end{array}
\right .,
\label{eq2}
\end{eqnarray}

\noindent
where $A = 7.049 556 277$, $B = 0.602 224 558 4$, and $a = 1.8$. The
repulsive three-body potential is also short-ranged, and is given by

\begin{eqnarray}
v_3({\bf r}_i,{\bf r}_j,{\bf r}_k) \equiv h(r_{ij}, r_{ik},\theta_{jik}) +  
h(r_{ij}, r_{jk},\theta_{ijk})  \nonumber \\
+  h(r_{ik}, r_{jk},\theta_{ikj}), 
\label{eq3}
\end{eqnarray}

\noindent
where $\theta_{jik}$ is the angle formed by the vectors ${\bf r}_{ij}$
and  ${\bf r}_{ik}$ and

\begin{eqnarray}
h(r_{ij}, r_{ik},\theta_{jik}) &= &\epsilon 
\lambda \exp[\frac{\gamma}{r_{ij}-a} + \frac{\gamma}{r_{ik}-a}] 
\left(\cos\theta_{jik}+\alpha \right)^{2} \nonumber \\
&&\times  H(a-r_{ij}) H(a-r_{ik})  ,
\label{eq4}
\end{eqnarray}

\noindent
where $\lambda =21.0, \gamma = 1.20$, and $H(x)$ is the  Heaviside
step function. The choice $\alpha = 1/3$ in $\left(\cos \theta_{jik} +
\alpha\right)^{2}$ favors a tetrahedral arrangement of atoms as found
in silicon. 

While the two-body interaction favors a close packed arrangement of
particles, the three-body interaction term favors an open structure
whose geometry depends on the parameter $\alpha$. The balance between
the two-body and three-body interactions, controlled by the parameter
$\lambda$ dictates the final preferred geometry of particle
arrangements \cite{Molinero}. The idea behind the modification of the
SW potential is to adjust the three-body interactions so that the
average coordination number is reduced, which in turn will shrink the
phase-coexistence curves and thus increase the region in the $T-\rho$
diagram in which gels can form, without the involvement of
phase-separation. Although similar in spirit to models, {\it e.g.} in
\cite{Russo,Bianchi}, in our model the number of bonds depends on
density and temperature, which is more similar in this respect to the
model studied in \cite{KobEuro}. 

By appropriately choosing the values of $\lambda$ and $\alpha$, it is
possible to obtain a system that has only a small coexistence region, 
with a network morphology whose average connectivity can be
tuned. We choose (among other possible choices) values $\lambda=10$ 
and $\alpha=1.49$ since these values avoid also the formation of ordered structures
(further details may be found in Ref\cite{shibu-thesis,EPAPS}).
In the following, all quantities are reported in reduced units for 
the modified Stillinger-Weber potential, i.e.~$\epsilon$, $\sigma$ 
and $m$ are the units of energy, length, and mass.

We have performed molecular dynamics (MD) simulations in the NVT
ensemble using $4000$ particles in a cubic box with periodic boundary
conditions.  We have used the constraint method\cite{Brown-Clarke}
for constant temperature simulation, where the equations of motion are
modified to maintain a constant kinetic energy.  The MD simulations for
the SW potential are computationally demanding due to presence of the
three-body interaction energies, where angles of all the triplets need
to be determined. Therefore, we have used an efficient method which allows
the calculation of the three-body interaction term by means of a two loop
summation\cite{Tiledesley,Biswas-Hamann,W-S1993,Makhov-Lewis,Frenkel-2loop}.
We have extended this approach to calculate the force due to three-body
interactions, as we describe in the Appendix. 
We have used a time step of $\Delta t=0.005$
in the MD simulations. The temperatures investigated are $T=5.0$, $3.0$, 
$1.0$, $0.30$, $0.20$, $0.10$, $0.09$, $0.08$, $0.07$, $0.06$, $0.05$, 
$0.04$, $0.03$, and $0.028$ at density $0.06$. Most of the data are averaged over
$5$ different samples. We have used $20$ to $30$ million MD steps for
equilibration and $30$ to $100$ million MD steps for data generation.

The Monte Carlo (MC) simulations have been performed with $512$ particles,
typically using a maximum step size of $0.085$. 
In the MC simulations, $2$ million MC steps were used for equilibration and $10$
to $20$ million MC steps for data generation.


\section{Static Properties}

\begin{figure}[t]
\includegraphics[scale=0.50,angle=0]{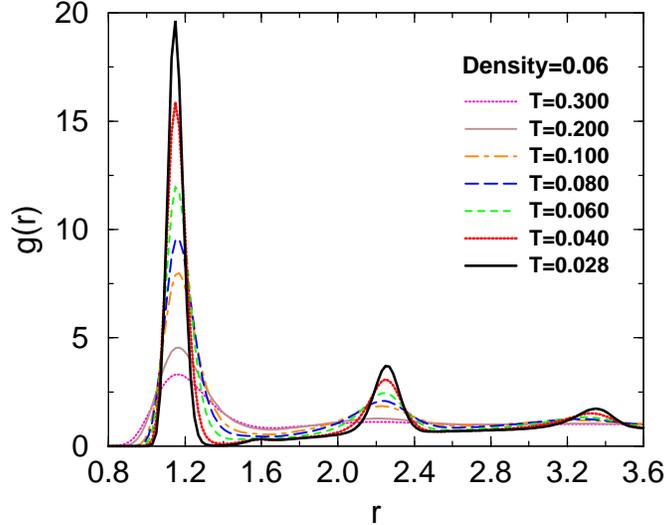}
\caption{The radial distribution function, $g(r)$, at density
$\rho=0.06$ for different temperatures. At high temperatures, $g(r)$
does not show much structure beyond the first peak. At low
temperatures $g(r)$ exhibits multiple peaks corresponding to well
defined separation distances for second, third, ... nearest neighbors.}
\label{gofr}
\end{figure}

Figure~\ref{gofr} shows the radial distribution function, $g(r)$,
for different temperatures at density $0.06$. At higher temperatures
$g(r)$ shows a single peak corresponding to the interparticle distance.
Additional small peaks appear at lower temperatures corresponding
to second and third nearest neighbor distances, similar to dense
liquids. However, here the $g(r)$ curves are characterized by relatively
sharp peaks, with the gaps in between having values of $g(r)$ close to the
ideal gas value of $1$ (except for low temperatures, between the first and
second peaks of the $g(r)$). These features arise from the fact that the
system is composed, at low temperatures, of long interconnected chains.

\begin{figure}[t]
\includegraphics[scale=0.47,angle=0]{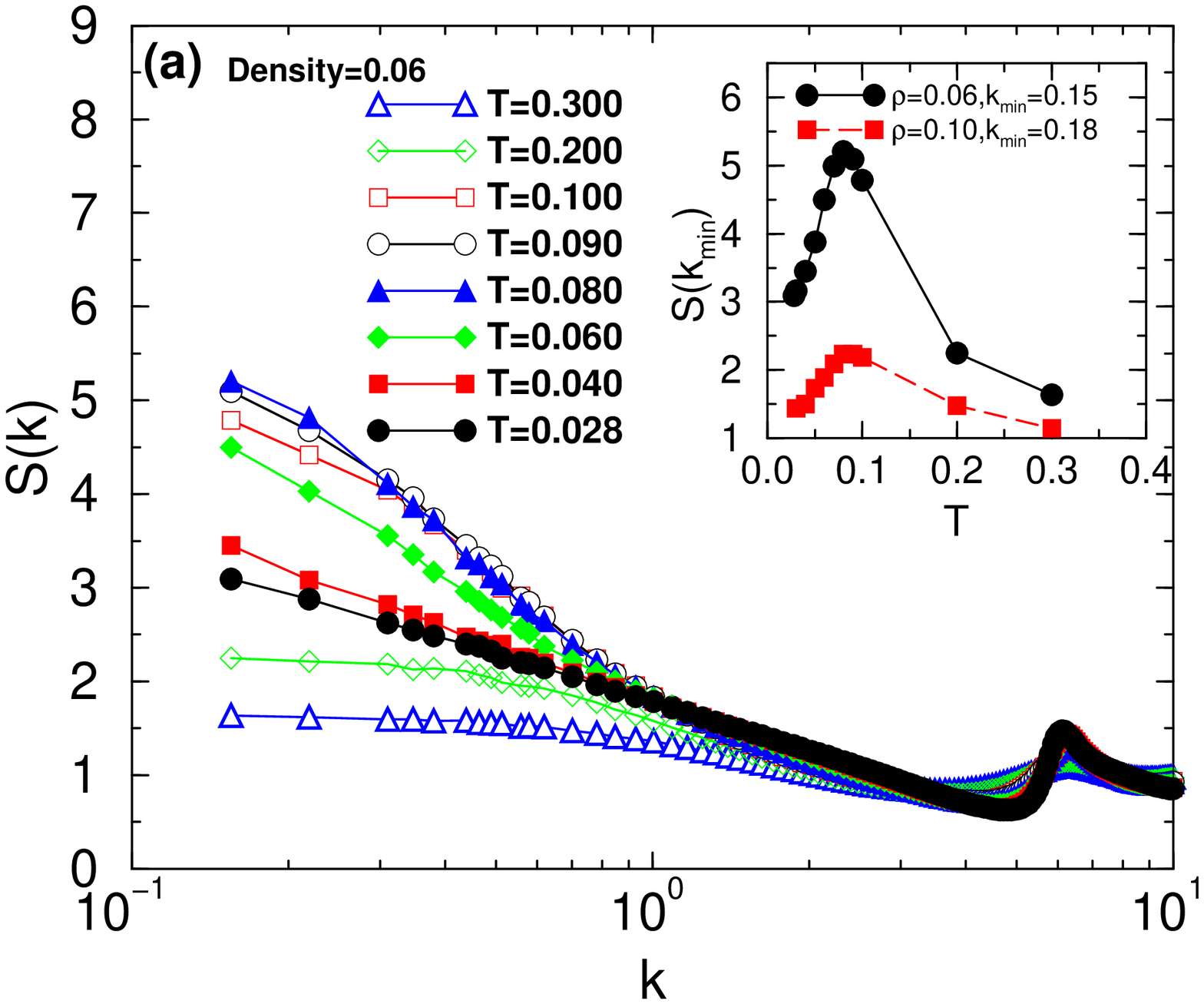}\vspace{3mm}
\includegraphics[scale=0.48,angle=0]{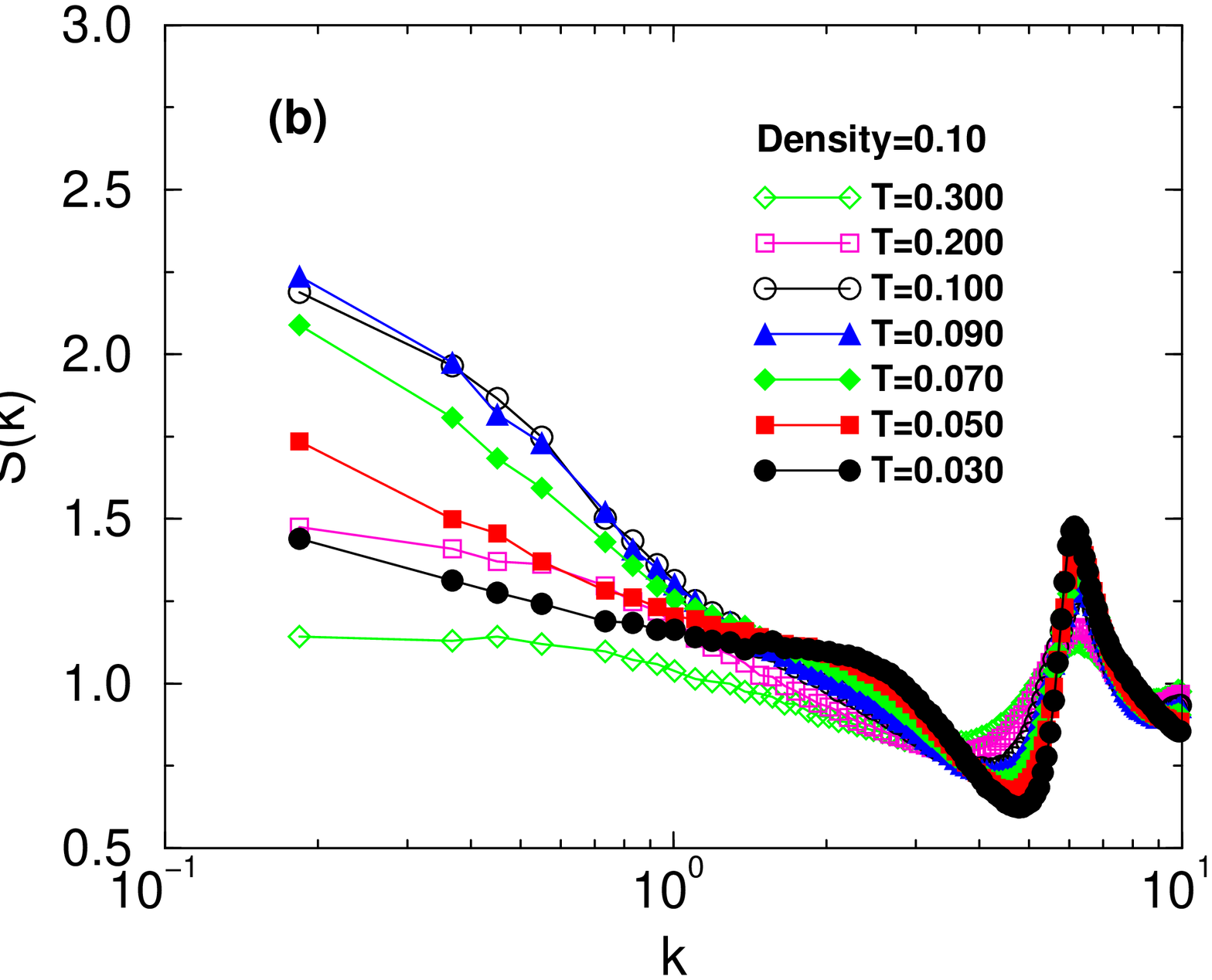}
\caption{The structure factor for  different temperatures at (a)
$\rho=0.06$ and (b) $\rho=0.10$. The $S(k)$ at the lowest wave-vector $k$
shows non-monotonic behavior in $T$ for both densities, as shown in the inset
of panel (a).  Temperatures below the maximum of $S(k_{min})$ are shown
in solid symbols while those above are shown in open symbols to clearly
indicate the non-monotonic behavior of $S(k)$.}
\label{Sk-k-rho-0.06-0.10-inset}
\end{figure}

In order to study the structure on large length scales, it is useful
to consider the structure factor, $S(k)$, defined as

\begin{equation}
 S({\bf k}) = \frac{1}{N} \sum_{j=1}^{N}  \sum_{l=1}^{N}   
\left< \exp(-\mathrm{i} {\bf k} \cdot ({\bf r}_j-{\bf r}_l)) \right>  \quad ,
\end{equation}

\noindent
where $N$ is the total number of particles and {\bf k} is the wave-vector.
Figure \ref{Sk-k-rho-0.06-0.10-inset} shows the $S(k)$ for a range of
temperatures at density $0.06$ and $0.10$.

At high temperatures, $S(k \rightarrow 0)$ has a value close to the
ideal gas value of $1$, but as the temperature is lowered, the $S(k)$
at small $k$ values gets substantially larger \cite{KobEuro}. Note that
at intermediate and small wave vectors, $S(k)$ shows a non-monotonic
dependence on temperature, displaying a maximum value at intermediate
temperatures (see inset in Fig.~\ref{Sk-k-rho-0.06-0.10-inset}a where
we show the $T$ dependence of $S(k_{min})$, i.e.~the value of $S(k)$
at the smallest wave-vector $k_{min}$ that is compatible with the
size of the simulation box). For $\rho = 0.06$, the peak occurs at
$T =0.08$, i.e.~well below the percolation transition at $T = 0.115$ \cite{EPAPS},
where one may naively have expected the maximum to lie.  The reason for
the shift away from the percolation point may be rationalized by noting
that in addition to percolation, which would give rise to a power-law in
$S(k)$, the structure of the system changes significantly because of the
formation of increasingly longer and stiffer linear chains. The latter
effect will, {\it via} the form factor of the chains, also give rise
to an increase in $S(k)$ at low $k$. The net effect is a shift of the
temperature at which $S(k)$ has the maximum value at low wave vectors
(see phase-diagram in Ref.\cite{EPAPS}). We have calculated $S(k)$ for system sizes
$N = 512, 4000$, and $8000$ for some state points in order to investigate
finite size effects and do not find any significant size effects in the
form of $S(k)$.

\begin{figure}[t]
\includegraphics[scale=0.48]{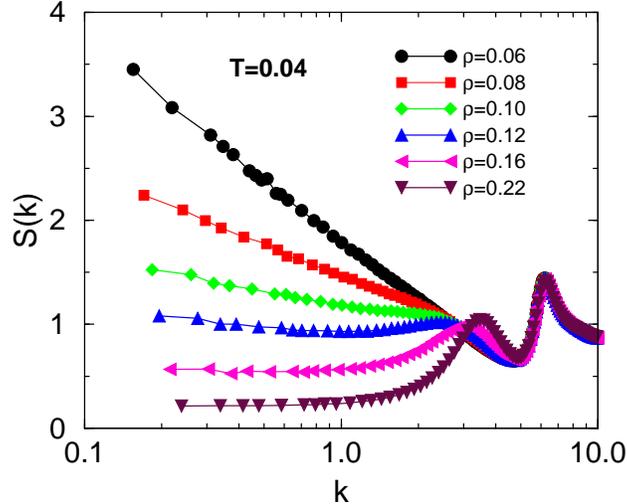}
\caption{The density dependence of $S(k)$ for $T=0.04$.  For  densities
above 0.1, the large values of $S(k)$ at low $k$ get suppressed, and a
finite $k$ peak develops around $k=3.0$.}
\label{Sk-rho-T-0.04}
\end{figure}

The density dependence of $S(k)$ at $T=0.04$ is shown in
Fig. \ref{Sk-rho-T-0.04}. At low densities, $\rho < 0.12$, the first
peak is at $k = 6.15$ corresponding to the nearest neighbor distance.
With an increase of the density, a peak around $k=3.0$ develops,
becoming pronounced at the highest density shown; the value of $S(k)$
at the smallest wave-vectors gets suppressed, indicating a reduction in
the compressibility and density fluctuations on long wavelengths.

\begin{figure}[b]
\includegraphics[scale=0.50,angle=0]{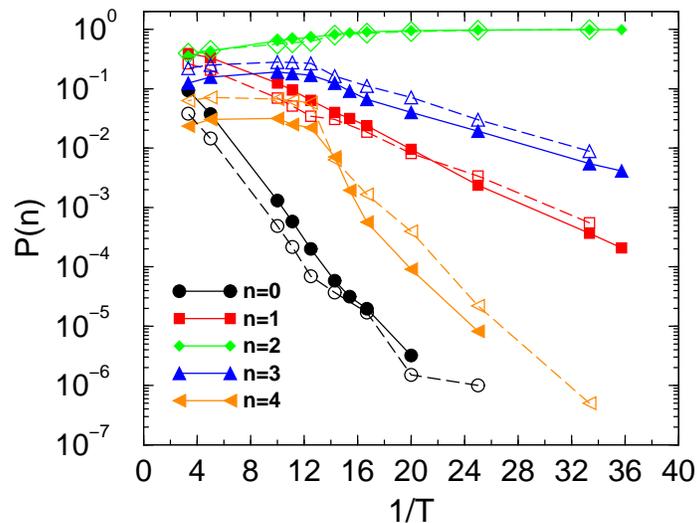}
\caption{The $T$-dependence of $P(n)$, the fraction of particles
having coordination number $n$, at density $0.06$ (solid symbols) and
$0.10$ (open symbols). At low temperatures most of the particles have
coordination $n=2$. With the decrease of temperature all coordinations
decrease except the one for $n=2$.}
\label{Pn-T}
\end{figure}

The local geometry of the system can be characterized by the
distribution of the coordination number $n$. Two particles are
considered to be neighbors if their distance is less than the ($T$ and
$\rho$-dependent) location of the first minimum of $g(r)$.  Figure
\ref{Pn-T} shows the temperature dependence of $P(n)$, the fraction of
particles having coordination number $n$, {\it vs.} $1/T$ for density
$\rho=0.06$ (filled symbols) and $\rho=0.10$ (open symbols). For $n =
0$ and 1, the $P(n)$ decreases rapidly and monotonically as $T$
decreases, whereas for $n=2$ the $P(n)$ increases monotonically with
decreasing $T$. These changes correspond to the formation of a network
of particles at the expense of isolated particles or dimers.
Interestingly for $n=3$ and 4 the $T-$dependence of $P(n)$ is
non-monotonic, displaying a maximum at the temperature at which
$S(k_{min})$ displays a maximum (see
Fig. \ref{Sk-k-rho-0.06-0.10-inset}). Above this temperature of
maximum $P(3)$ and $P(4)$, the structural change upon lowering $T$
arises from a growth in the number of bonds leading to percolation. In
contrast, at temperatures below the location of the maximum, the
system forms increasingly longer linear chains at the expense of
cross-links with larger coordination. In contrast, in the case of the
gel-forming model in \cite{KobEuro}, P(3) increases with lowering of T
and saturates at very low temperatures. The fraction of three-fold
coordinated particles, $P(3)$, increases strongly with density (data
not shown). Thus at higher density there are more anchor particles and
consequently the lengths of the linear chain segments are smaller
compared to that at low density.

We have also calculated the cluster size distribution for density
$0.06$ \cite{EPAPS}. Even for temperatures near the percolation transition, we do
not find a distinguishable power-law regime in the cluster size
distribution, in contrast to the findings for other gel forming
systems for which a power-law dependence corresponding to random
percolation has been observed
\cite{KobEuro,Sciortino-ns,Coniglio-ns}. We believe this to arise from
the fact that while the (exponential) cluster size distribution is
obtained from a random aggregation process at high temperatures, the
percolation transition is influenced by the emergence of increasingly
long linear chains.

\begin{figure}[t]
\includegraphics[scale=0.45,angle=0]{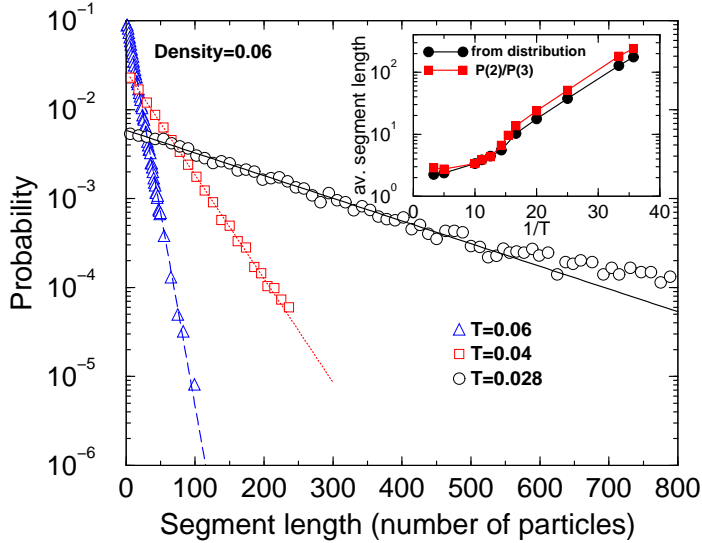}
\caption{The segment length distribution at density $0.06$ for different
temperatures, which exhibit exponential behavior. The lines are
exponential fits to the data. Inset: The average segment length {\it vs.}
$1/T$ from exponential fits in the main panel, and the estimate using
the ratio of number of two and three coordinated particles, $P(2)/P(3)$.}
\label{seg-N}
\end{figure}

The distribution of segment lengths shows an exponential behavior (see
Fig.~\ref{seg-N}), with an average segment length that increases with
decreasing temperature, as can also be inferred from the $T$-dependence
of $P(2)$ and $P(3)$ in Fig.~\ref{Pn-T}. In the inset of Fig.~\ref{seg-N}
we show the average segment length obtained from an exponential fit to
the distribution of segment lengths, as well as the ratio $P(2)/P(3)$,
which provides an estimate of the average segment length. These two are in
good agreement. For other densities also, the segment length distribution
exhibits exponential behavior and the average segment length decreases
with an increase of density.

The squared end-to-end displacement $\langle \delta^2(N_c)\rangle$ of chain
segments shows a quadratic behavior for small chain lengths $N_c$ and
is linear for large $N_c$~\cite{shibu-thesis}.  The persistence length
can be defined as the maximum length of a chain for which $\langle
\delta^2(N_c)\rangle \sim N_c^2$ holds. At $T = 0.028$ and density $0.06$,
the persistence length is 15 and it decreases with increasing temperature
or density.

\section{Dynamics}

\begin{figure}[t]
\includegraphics[scale=0.45,angle=0]{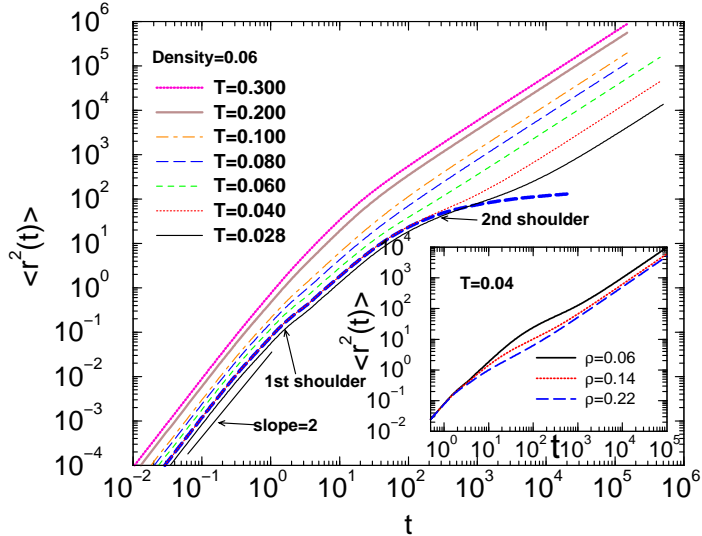}
\caption{Mean squared displacements for different temperatures at
$\rho=0.06$ in a log-log plot. The location of the $1^{st}$ and $2^{nd}$
shoulders, seen for low $T$ and discussed in the text, are marked. The
bold dashed line shows the MSD from constrained simulations at $T=0.04$
(see text) which saturates at the
second shoulder. The inset shows the MSD at the lowest $T$ for different
densities.}
\label{MSD}
\end{figure}

In order to study the dynamics of the system we first consider the mean
squared displacement (MSD) of particles,  defined as

\begin{equation}
\langle r^2(t)\rangle \equiv \frac{1}{N} \sum_{i=1}^{N}  \langle 
\Big|{\bf r}_{i}(t)-{\bf r}_{i}(0)\Big|^2 \rangle, 
\end{equation} 

\noindent
which is shown in Fig.~\ref{MSD} for different $T$ at density $0.06$.

At high temperatures, a smooth crossover is seen between the short
time ballistic and long time diffusive regimes. If the temperature
is lowered, one sees the emergence of incipient plateaus, as seen in
dense supercooled liquids.  However, unlike dense liquids, one observes
the presence of {\it two} ``shoulders'', rather than one \cite{KobPRL}. The first
appears at a length of $\sim 0.4 \sigma$, analogous to what is seen in
dense liquids.  The second shoulder appears at $\sim 7 \sigma$,
indicating that transient localization occurs here at a much larger
length scale compared to dense liquids. Although large, this length
is significantly smaller than the average segment length of the chains
(around $100$ at the lowest $T$; see inset of Fig. \ref{seg-N}).  Thus,
we conclude that the amplitude of floppy motions contributing to the
MSD at the second shoulder is much smaller than the length of the chains.

The inset in Fig. \ref{MSD} shows the evolution of the MSD at $T=0.04$
with density. As density increases, the displacement corresponding to
the second shoulder decreases and also it becomes less pronounced since
the segment length between anchor particles decreases.

Also included in the main panel of Fig. \ref{MSD} is the MSD as obtained
from constrained MD simulations at $T=0.04$, i.e.~a dynamics in which
bonds are prevented from breaking or forming \cite{footnote1}.  As we see
in the figure, the MSD from constrained MD simulations saturates at the
second shoulder, indicating that the caging associated with the second
shoulder is that of chain segments executing floppy motion between two
anchor particles, whose magnitude is constrained by the network topology.

The non-Gaussian parameter has been studied extensively \cite{kobhet} in
the context of supercooled liquids as a way of characterizing dynamical
heterogeneities. The non-Gaussian parameter $\alpha_2(t)$ is defined
as\cite{Boon,odagaki_hiwatari_91}

\begin{equation}
\alpha_2(t) \equiv \frac{3\langle r^4(t)\rangle }{5\langle r^2(t)\rangle ^2} -1  
\quad ,
\end{equation}

\noindent
where 
\begin{equation}
\langle r^4(t)\rangle  \equiv \frac{1}{N} 
\langle  \sum_{i=1}^{N} \Big|{\bf r}_{i}(t)-{\bf r}_{i}(0)\Big|^4\rangle  \quad .
\end{equation} 

\begin{figure}[h]
\includegraphics[scale=0.50,angle=0]{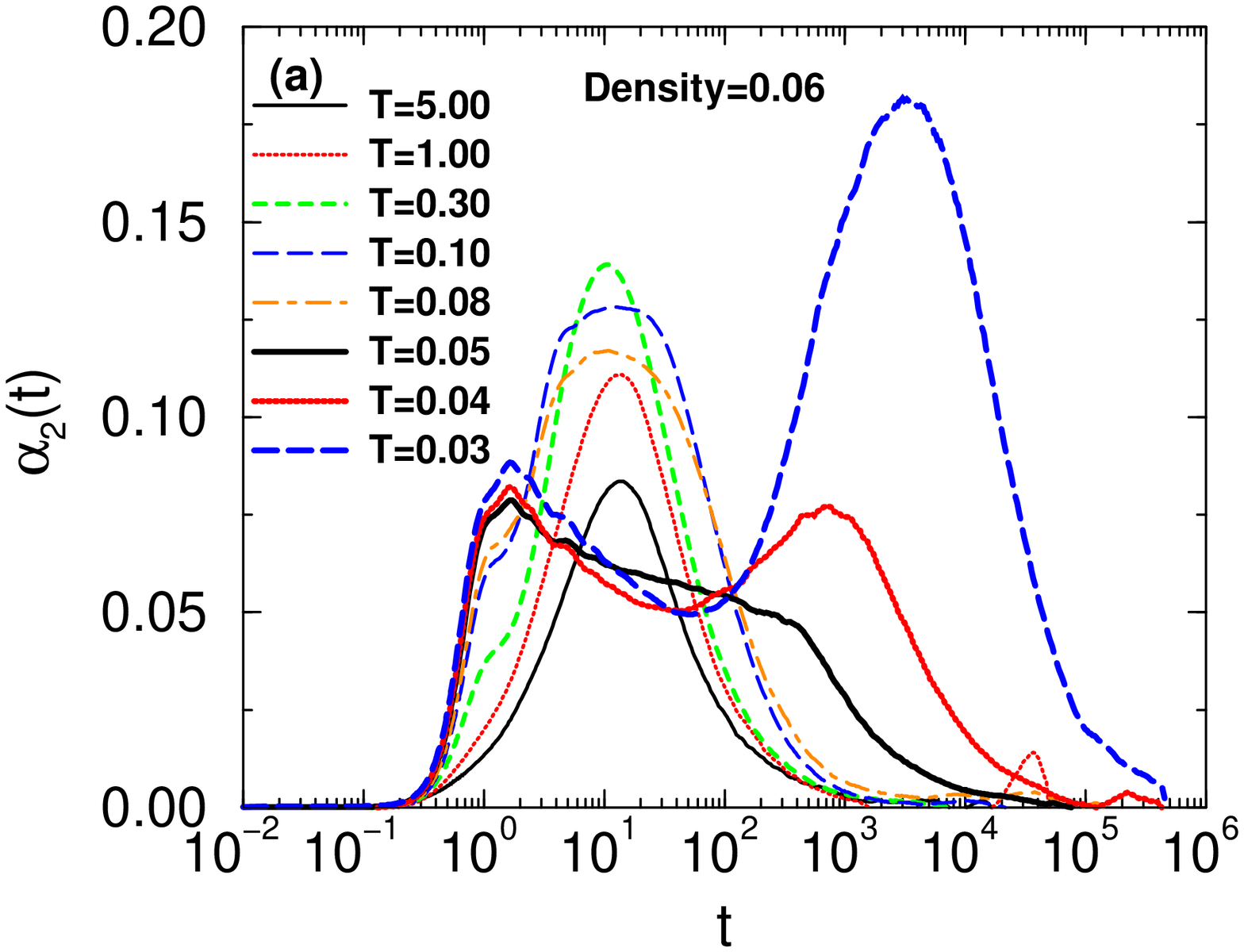}
\includegraphics[scale=0.50,angle=0]{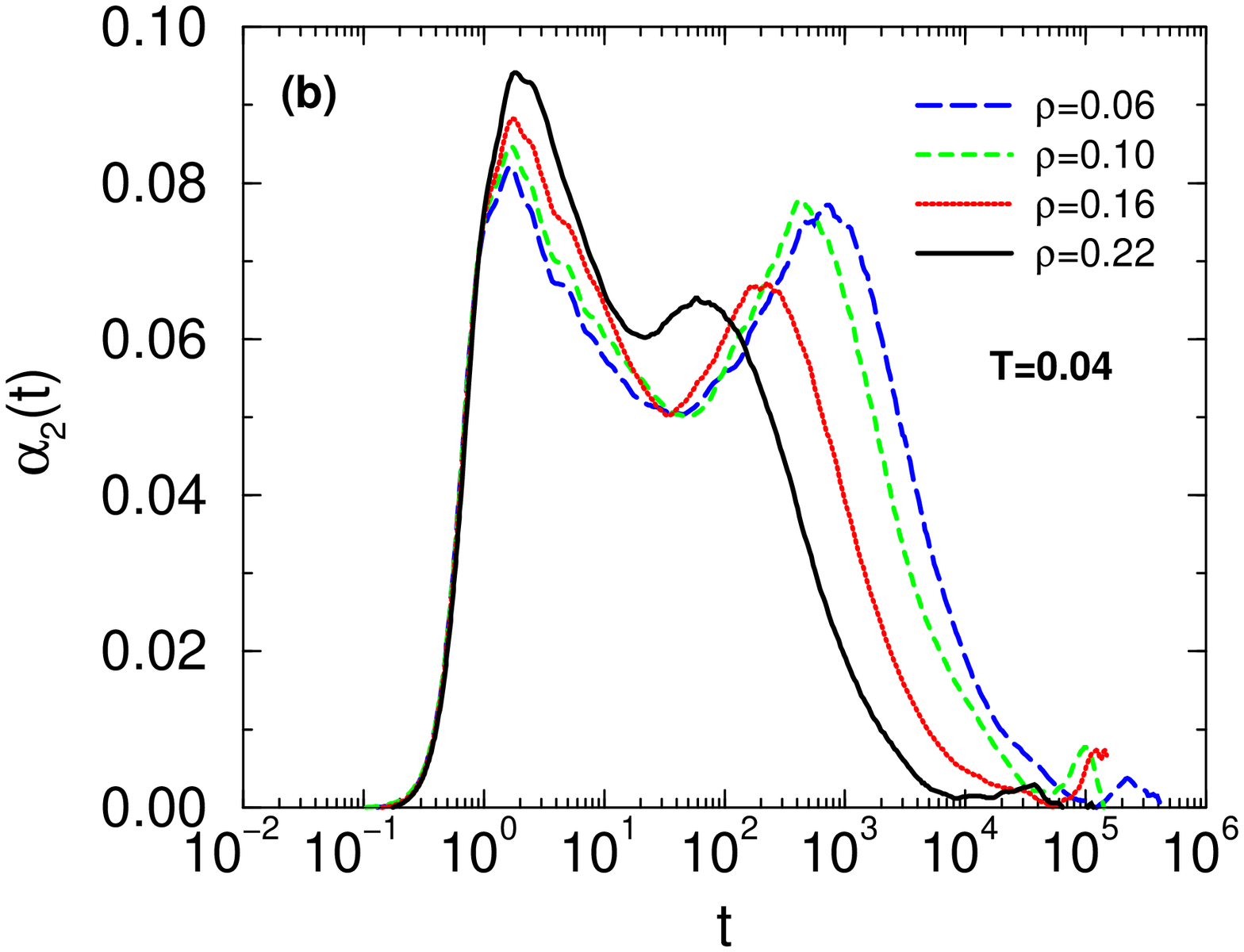}
\caption{The non-Gaussian parameter $\alpha_2(t)$ for (a) different $T$
at density $0.06$ and (b) different densities at $T=0.04$.}
\label{alpha2}
\end{figure}

Figure \ref{alpha2}a shows $\alpha_2(t)$ for different $T$
at density $0.06$.  At high temperatures, a single peak in $\alpha_2(t)$
is observed at around $t=10$, a time that corresponds to the crossover in the MSD
from ballistic to diffusive behavior.  At intermediate temperatures, an
additional peak at $t \approx 1$ emerges, which is the time at which the
first shoulder in the MSD is observed. Its origin is likely the heterogeneous 
dynamics related to the fact that different particles can have quite different type
of cages since they can be two or three-fold coordinated.
If $T$ is decreased even further,
the location of the peak at longer times rapidly
shifts to larger times, and its height increases with decreasing $T$.
This peak corresponds to the second shoulder seen in the MSD. We note that
the height of this peak is, even at the lowest temperatures, less
than 0.2, a value that is significantly smaller than the ones that are
found in {\it dense} glass-forming systems, such as, e.g.~Lennard-Jones
mixtures~\cite{kob_andersen_95a} for which $\alpha_2(t_{max}) \approx
1.5$. This difference is probably due to the fact that in the present
system there is, at low temperatures, not that much variance in the
relaxation dynamics of the individual particles since the main relaxation
process is a cutting of a chain and a subsequent reconnection of the
loose ends to the rest of the network. This process depends only very
weakly on the particle considered and hence the $\alpha_2(t)$ is small.

In Fig.~\ref{alpha2}b we show the density dependence of the non-Gaussian
parameter for $T=0.04$. The position of the first peak, around $t =
1$, is independent of density whereas the second peak position moves
to larger times with decreasing density, in agreement with the density
dependence of the MSD {\it on the time scale of the second shoulder}.

To study the life time of the bonds, we calculate the bond correlation
function $\phi_B(t)$ defined as\cite{Sciortino,Starr1,Starr2}

\begin{equation}
\phi_{B}(t)=\frac{\left<\sum_{i<j}\delta n_{ij}(t)\delta 
n_{ij}(0)\right>}{\left<\sum_{i<j}\delta n_{ij}(0) \delta  n_{ij}(0)\right>}  \quad ,
\end{equation}

\noindent
where

\begin{eqnarray} 
\delta  n_{ij}(t)&=&n_{ij}(t) -\langle n\rangle  \nonumber \\  
\mbox{and }n_{ij}&=& 1 - H(r_{ij}-r_{cut}) \quad .  
\end{eqnarray}

\begin{figure}[h]
\includegraphics[scale=0.50,angle=0]{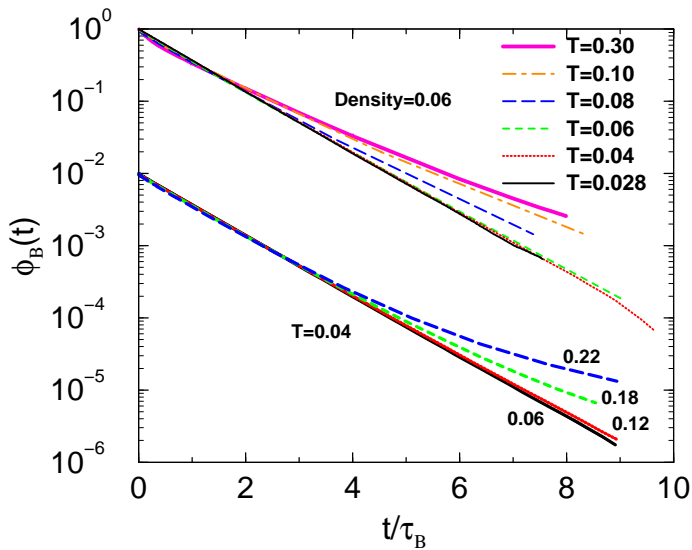}
\caption{The bond correlation function $\phi_B(t)$ as a function
of $t/\tau_B$ at different
temperatures and density $0.06$ (upper curves ) and at different densities
for $T=0.04$ (lower curves). The latter set of curves has been shifted
down by a factor of $100$ for the sake of clarity. $\tau_B$ is the average 
bond life time. }
\label{phiBt}
\end{figure}

\noindent
$H(x)$ is the Heaviside function and $r_{cut}$ is the bond length defined
as the distance at which the first minimum of $g(r)$ occurs. $n_{ij}$
is $1$ when a bond is present and $0$ otherwise.  Hence, $\phi_B(t)$
counts the fraction of bonds found at $t=0$ that survive after a time
$t$, independent of any breaking and reforming at intervening times.
Figure \ref{phiBt} shows the bond correlation function at density $0.06$
for different temperatures (upper curves) and at $T=0.04$ for different
densities (lower curves). (For the sake of clarity, the lower curves have
been shifted down by a factor of $100$.) In order to recognize better
the $T$ and $\rho-$dependence of this correlation function we plot it as
a function of $t/\tau_B$, where $\tau_B$ is the average bond life time
as determined from the area under $\phi_B(t)$.  We see that the bond
correlation functions decay exponentially with time at low $T$, but that the
decay is slower than exponential at intermediate temperatures.  At high
$T$ the decay is again exponential (data not shown). Similarly,
at $T=0.04$, the decay behavior is exponential at low densities and
stretched exponential at higher densities. In both cases, the stretched
exponential behavior appears to be associated with an increase of disorder
in the local environment of the bonds.

\begin{figure}[b]
\includegraphics[scale=0.50,angle=0]{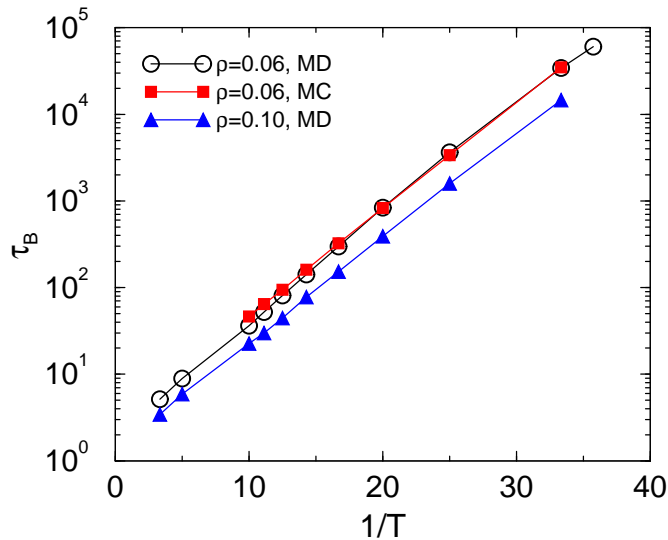}
\caption{The bond lifetime $\tau_B$ {\it vs.} inverse $T$ at density
$0.06$ and $0.10$ from MD simulations and from MC
simulations for density $0.06$.}
\label{tauB-T}
\end{figure}

Figure~\ref{tauB-T} shows the $T$-dependence of the bond life time,
$\tau_B$, for density $0.06$ and $0.10$ from MD simulations as well as
the $\tau_B$ from MC simulations at density $0.06$. We recognize from
the figure that the bond life times from MD and MC simulations compare
well with each other after scaling the MC times (expressed in number of
sweeps) by a single multiplicative factor of 100. This shows
that the relaxation dynamics does not depend on the details of the microscopic 
dynamics. $\phi_B(t)$ exhibits
an Arrhenius behavior with an activation energy of $0.28$.

One of the most useful quantities for the study of dynamics is the
intermediate scattering function $F(k,t)$ defined as \cite{Hansen-McDonald}

\begin{equation}
F(k,t) = \frac{1}{N} \sum_{l=1}^{N}  \sum_{j=1}^{N}  
\langle \exp[-\mathrm{i} {\bf k} \cdot ({\bf r}_l(t)-{\bf r}_j(0))]\rangle.  
\end{equation}

\noindent
$F(k,t)$ provides information on the collective dynamics of the
system. The {\it self} intermediate scattering function, $F_s(k,t)$,
which reveals information about single particle motion, is obtained by
restricting the double summation above to $l = j$:

\begin{equation}
F_s(k,t) = \frac{1}{N} \sum_{j=1}^{N}   
\langle \exp[-\mathrm{i} {\bf k} \cdot ({\bf r}_j(t)-{\bf r}_j(0))] \rangle. 
\end{equation}

\begin{figure}[t]
\includegraphics[scale=0.50,angle=0]{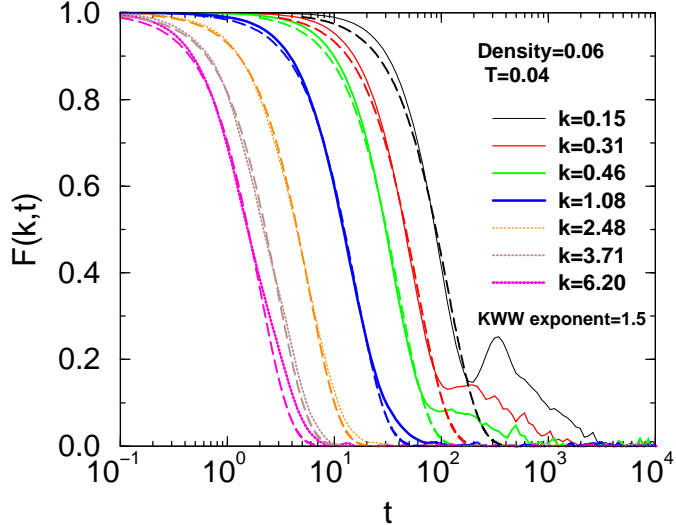}
\caption{The collective intermediate scattering function $F(k,t)$ at
$\rho=0.06$ and $T=0.04$ for different wave-vectors $k$. The KWW fit
to $F(k,t)$ at intermediate times with an exponent $\beta=3/2$ is also
included (dashed lines).}
\label{Fkt-t-fit-1.50}
\end{figure}

Figure~\ref{Fkt-t-fit-1.50} shows $F(k,t)$ {\it vs.} time $t$ for
different wave-vectors $k$ at $T=0.04$ and density $\rho=0.06$. From
this graph we recognize that $F(k,t)$ shows three regimes: At very
short times the function is quadratic in $t$, since we have a Newtonian
dynamics. At long times the function shows at small $k$ an oscillatory
behavior which is related to the usual acoustic sound modes. The
most remarkable feature is seen at intermediate times where we find
that $F(k,t)$ shows a {\it compressed exponential} decay, $F(k,t)
= \exp(-(t/\tau)^\beta)$, with an exponent $\beta \approx 3/2$ for all
$k$-values shown. (For $k$ values that are even larger than the ones shown
in Fig.~\ref{Fkt-t-fit-1.50},
one finds that $F(k,t)$ decays ballistically, i.e.~$\beta \approx 2$,
see Ref.~\cite{shibuPRL}.) Also included in the graph are fits with
a Kohlrausch-Williams-Watts (KWW) function with an exponent of $3/2$.
Such compressed exponential relaxation, i.e.~$\beta > 1$, has been observed
in experiments on non-equilibrium colloidal gels and it has been argued
that this compression is related to the stress inhomogeneities caused by
the shrinking of the gels\cite{Cipelletti,pitard}.  In the present case,
the system is in equilibrium and the compressed exponential behavior
is due to the floppy motion of the non-restructuring network present in
the gel-former\cite{shibuPRL}, as we will show below.

It is also of interest to note that even at this low temperature there
is no sign of the two-step relaxation observed in dense glass-forming
liquids (see, e.g., Ref.~\cite{kob_andersen_95b}). Thus we have here an
example of a system whose relaxation dynamics is glassy (non-exponential,
strong $T-$dependence, complex $k-$dependence,...) but with correlation
functions that do not show the caging regime at intermediate times
(at least in the $k$ and $T-$range accessible in this simulation).

\begin{figure}[b]
\includegraphics[scale=0.50,angle=0]{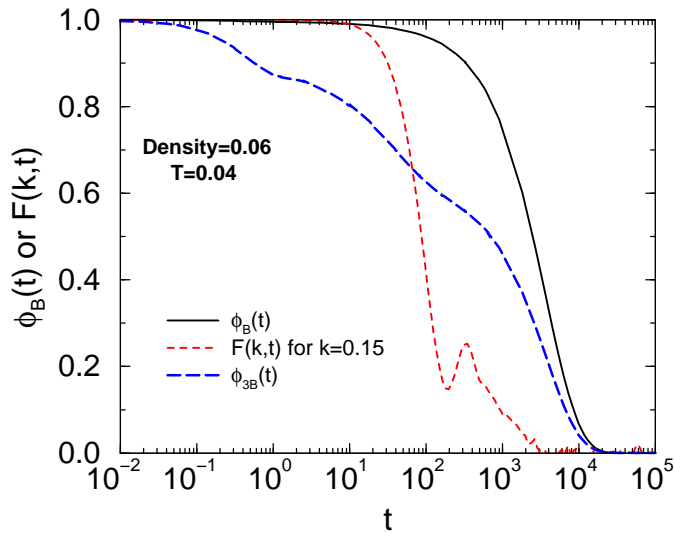}
\caption{The collective intermediate scattering function $F(k,t)$
and the bond correlation function $\phi_B(t)$ at density $0.06$ and
$T=0.04$. Also shown is the bond correlation function for particles
which are initially three-fold coordinated.}
\label{Fktwithbondlifetime}
\end{figure}

In order to get a better understanding of the origin of the compressed
exponential it is useful to compare the time dependence of $F(k,t)$ with
the one of the bond correlation function $\phi_B(t)$. This is done in
Fig.~\ref{Fktwithbondlifetime}, where we show $F(k,t)$ for the smallest
wave-vector as well as the bond correlation function for $T = 0.04$,
$\rho = 0.06$. Also included is the bond correlation function defined
only for particles that initially have three-fold coordination. It is
seen that $F(k,t)$ decays with a compressed exponential form on a time
scale that is much shorter than the bond life time for all particles, and
also in which a substantial fraction of bonds associated with initially
three-coordinated particles are still intact. This observation suggests
that the compressed exponential relaxation seen is due to the dynamics
of the gel network on time scales when bond breaking is not very relevant.

\begin{figure}[t]
\includegraphics[scale=0.50,angle=0]{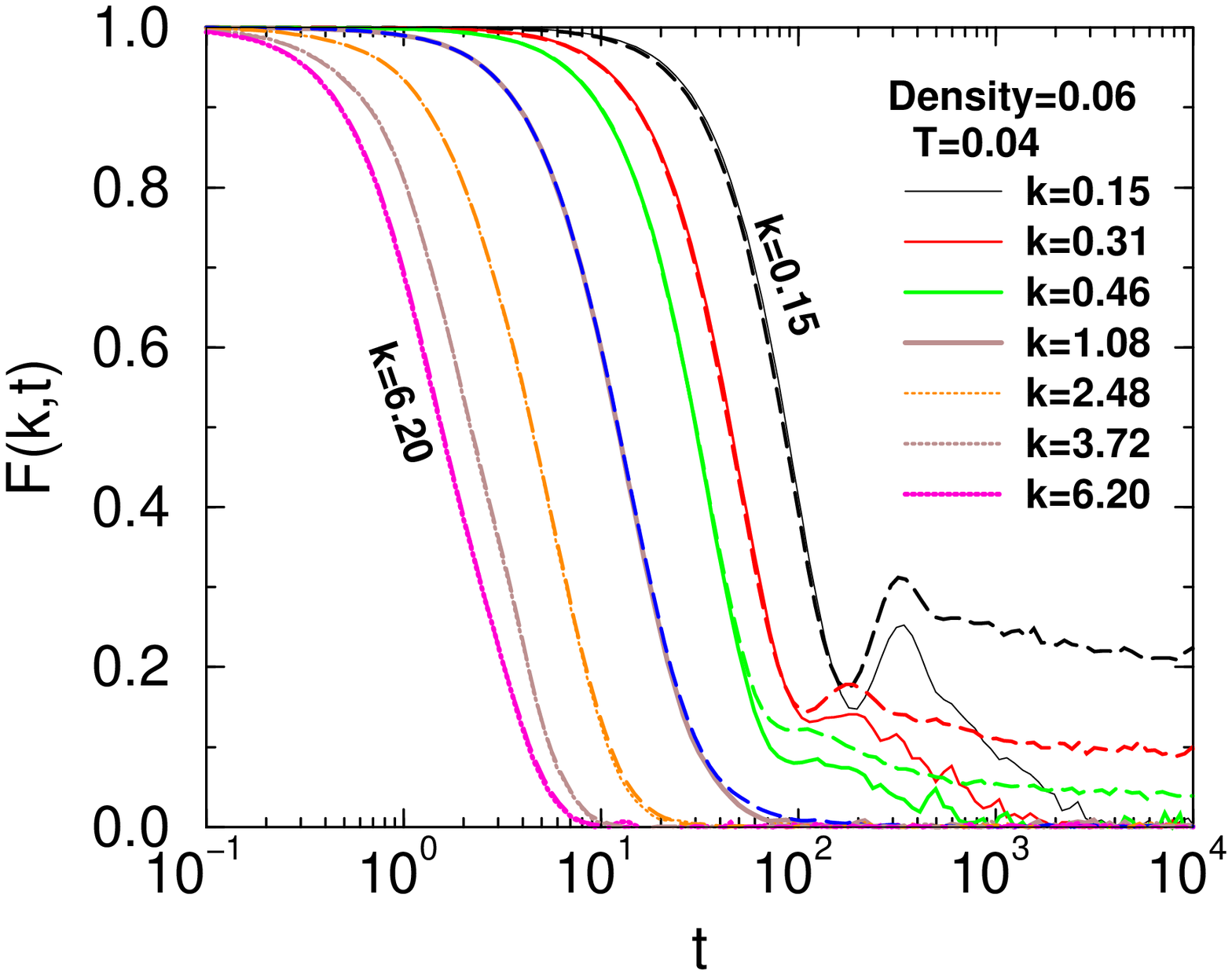}
\caption{The collective intermediate scattering function $F(k,t)$ for
different wave-vectors $k$ at density $\rho=0.06$ and $T=0.04$ from MD
(solid lines) and constrained MD simulation (dashed lines).}
\label{constraintMD}
\end{figure}

To confirm this, we have performed constrained MD simulations, i.e.~a
dynamics in which bonds are prevented from breaking or forming
\cite{footnote1}. Figure~\ref{constraintMD} shows $F(k,t)$ with and
without the presence of this constraint potential. The two $F(k,t)$
curves are essentially identical in the time regime where compressed
exponential relaxation is seen, but deviate from each other at later
times, with the constrained MD curve saturating to a finite value. The
data shown in Fig. \ref{constraintMD} clearly demonstrates that the
origin of the compressed exponential relaxation is in the vibrational
dynamics of the gel network.

\begin{figure}[b]
\includegraphics[scale=0.50,angle=0]{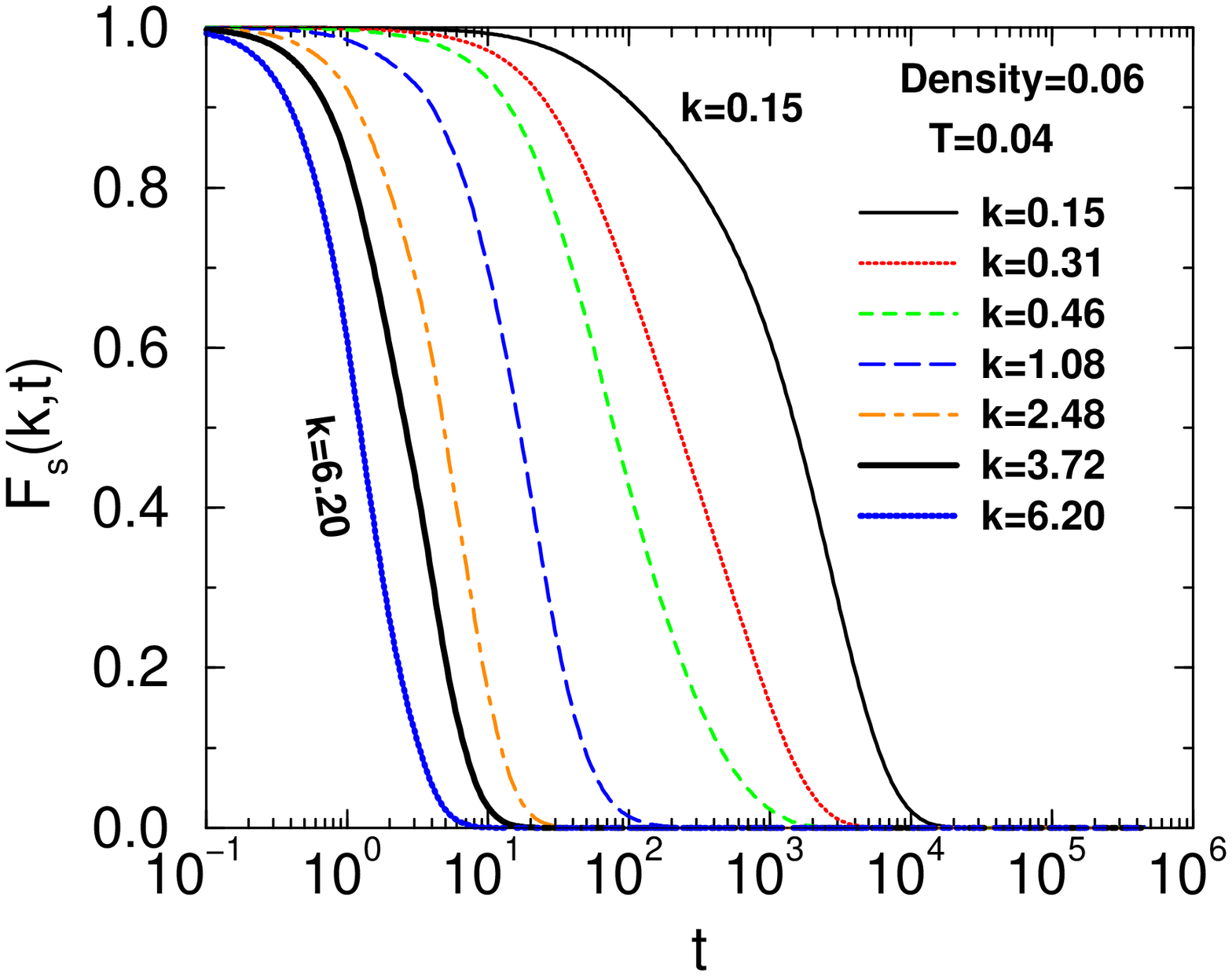}
\caption{The self intermediate scattering function $F_s(k,t)$ from MD
simulations  for $\rho=0.06$ and  $T=0.04$ for different wave-vectors
$k$.}
\label{Fskt-MD}
\end{figure}

Figure \ref{Fskt-MD} shows the time dependence of the self intermediate
scattering function $F_s(k,t)$ at density $0.06$ and temperature
$T=0.04$ for different wave-vectors. At long times, the decay is close to
exponential for small wave vectors and faster than exponential at large
wave vectors~\cite{shibuPRL}. A close inspection of Fig.~\ref{Fskt-MD}
reveals that the stretching exponent varies non-monotonically, and is
smaller at intermediate wave vectors. This behavior holds at other low
temperatures as well \cite{shibuPRL}. A comparison with the behavior of
$F(k,t)$ shows that these two functions indeed reveal very different
aspects of the dynamics of the gel former.

\begin{figure}[t]
\includegraphics[scale=0.50,angle=0]{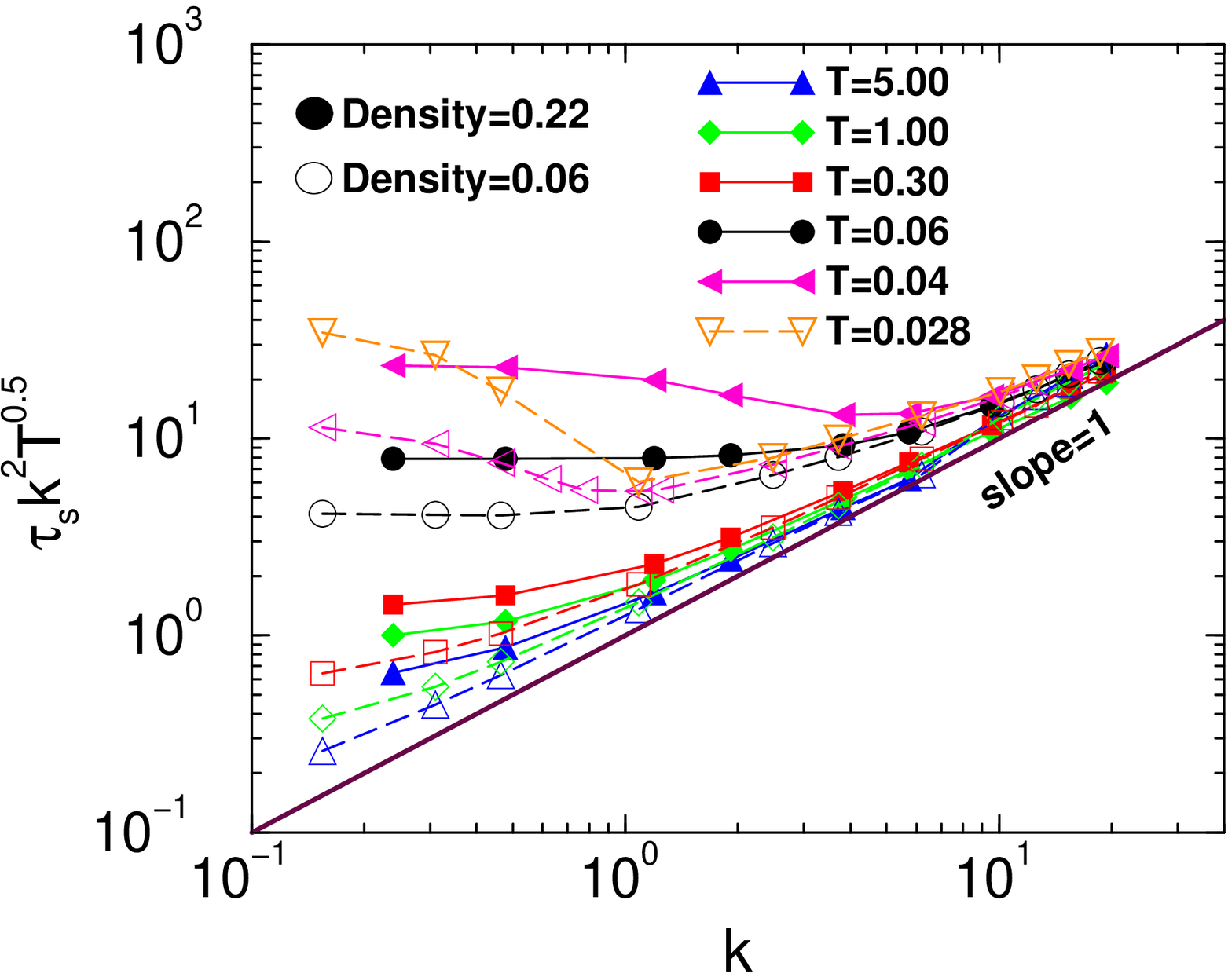}
\caption{The $k-$dependence of   $\tau_s $ multiplied by $k^2$, where
$\tau_s$ has been obtained from the area of $F_s(k,t)$, at different $T$
at $\rho=0.22$ (filled symbols) and $\rho=0.06$ (open symbols).}
\label{tausk2-k-rho-0.22-0.06}
\end{figure}

In Fig. \ref{tausk2-k-rho-0.22-0.06} we show the relaxation time $\tau_s$
that have been calculated from the area under $F_s(k,t)$ {\it vs.}  time
for different wave vectors at densities $0.06$ (open symbols) and $0.22$
(filled symbols). (In order to compensate the trivial $1/T^{0.5}$ dependence 
at large $k$, we multiply the data by $T^{0.5}$.) 
At high wave-vectors the linear behavior of $\tau_s k^2$
indicates the ballistic character of the dynamics, consistent with the
gaussian behavior of $F_s(k,t)$ shown in Fig. \ref{Fskt-MD} and the value
for the KWW exponent which is around 2.0 (see Ref.~\cite{shibuPRL}). For
intermediate temperatures, we observe a crossover to a regime where
$\tau_s k^2$ is roughly constant at small wave vectors, seemingly
indicating a diffusive dynamics. However, the $F_s(k,t)$ curves shown in
Fig. \ref{Fskt-MD} reveal that the dynamics is more complex than a simple
diffusive behavior in these cases. Note that the location in $k$ at which
this crossover is observed depends strongly on temperature and density,
which indicates the complex dependence of the relaxation dynamics on
$\rho$ and $T$. This conclusion is also supported by the observation that
at the lowest temperature investigated the mentioned plateau at low $k$
disappears (see Fig.~\ref{tausk2-k-rho-0.22-0.06}) since the stiffening of
the network pushes the hydrodynamic regime to even lower values of $k$,
in agreement with previous results~\cite{KobPRL}.

\begin{figure}[b]
\includegraphics[scale=0.50,angle=0]{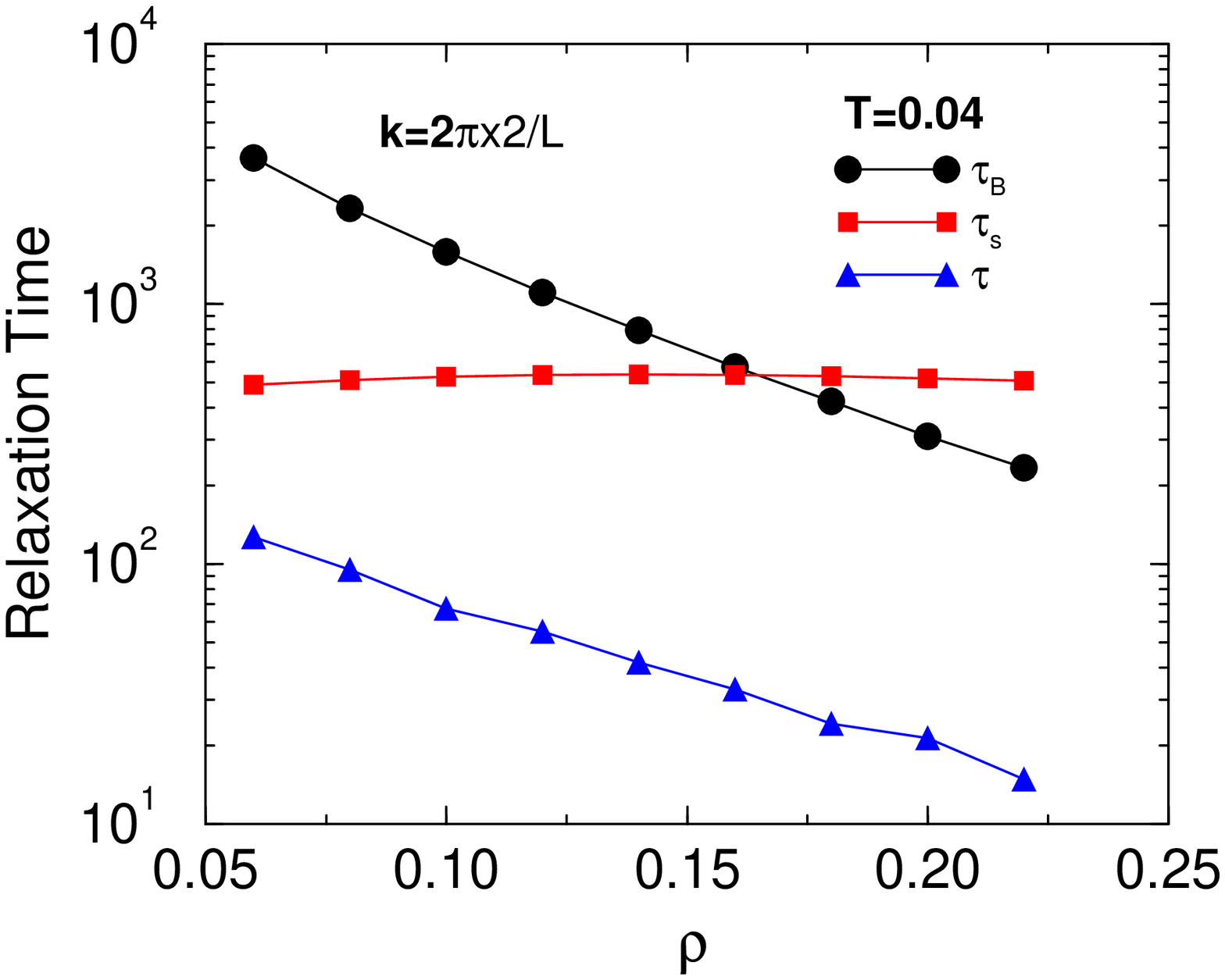}
\caption{The density dependence of the bond life time $\tau_{B}$ and
relaxation times $\tau$ and $\tau_s$ obtained from the area of the
coherent and incoherent intermediate scattering functions for $T=0.04$
and wave-vector $k= {2 \pi \over L} \times 2 L$, where $L$ is the size
of the simulation box.}
\label{tauB-tau-taus-rho-T-0.04-k-0.31}
\end{figure}

The relaxation times ($\tau$ and $\tau_s$) and $\tau_B$ at $T=0.04$
and wave-vector $k=2 \pi \times 2/L$, i.e.~ the second smallest
wave-vector compatible with the size $L$ of the simulation box, is shown
in Fig. \ref{tauB-tau-taus-rho-T-0.04-k-0.31}. The relaxation times
$\tau_s$, $\tau$ have been obtained from the area under $F_s(k,t)$,
$F(k,t)$ {\it vs.} time curves. We recognize that $\tau_B$ as well as
$\tau$ decrease with increasing density. The reason for this is that with
increasing $\rho$ the number of short lived three and higher coordinated
particles increases, with the consequences that: (i) $\tau_B$ of the
system decreases and (ii) the network structure can reconstruct more
easily, in turn leading to a faster decay of $F(k,t)$, and a decrease
of $\tau$.

On the other hand, $\tau_s$ is nearly independent of density. This is a
consequence of the fact that the MSD is roughly the same for all densities
in the diffusive regime (see Fig. \ref{MSD}). However, at larger
wave-vectors, $\tau_s$ shows a more significant dependence on density
(data not shown) \cite{shibu-thesis}, consistent with significant changes
in the MSD near the second shoulder.

Since in real experiments the relaxation dynamics of the gels is given
by a Brownian dynamics instead of the Newtonian dynamics studied here,
it is of interest to investigate which properties of the relaxation
dynamics depend on the microscopic dynamics. In the final part of this
paper we therefore compare $F(k,t)$ and $F_s(k,t)$ obtained from MD with
those obtained from MC.

\begin{figure}[h]
\includegraphics[scale=0.50,angle=0]{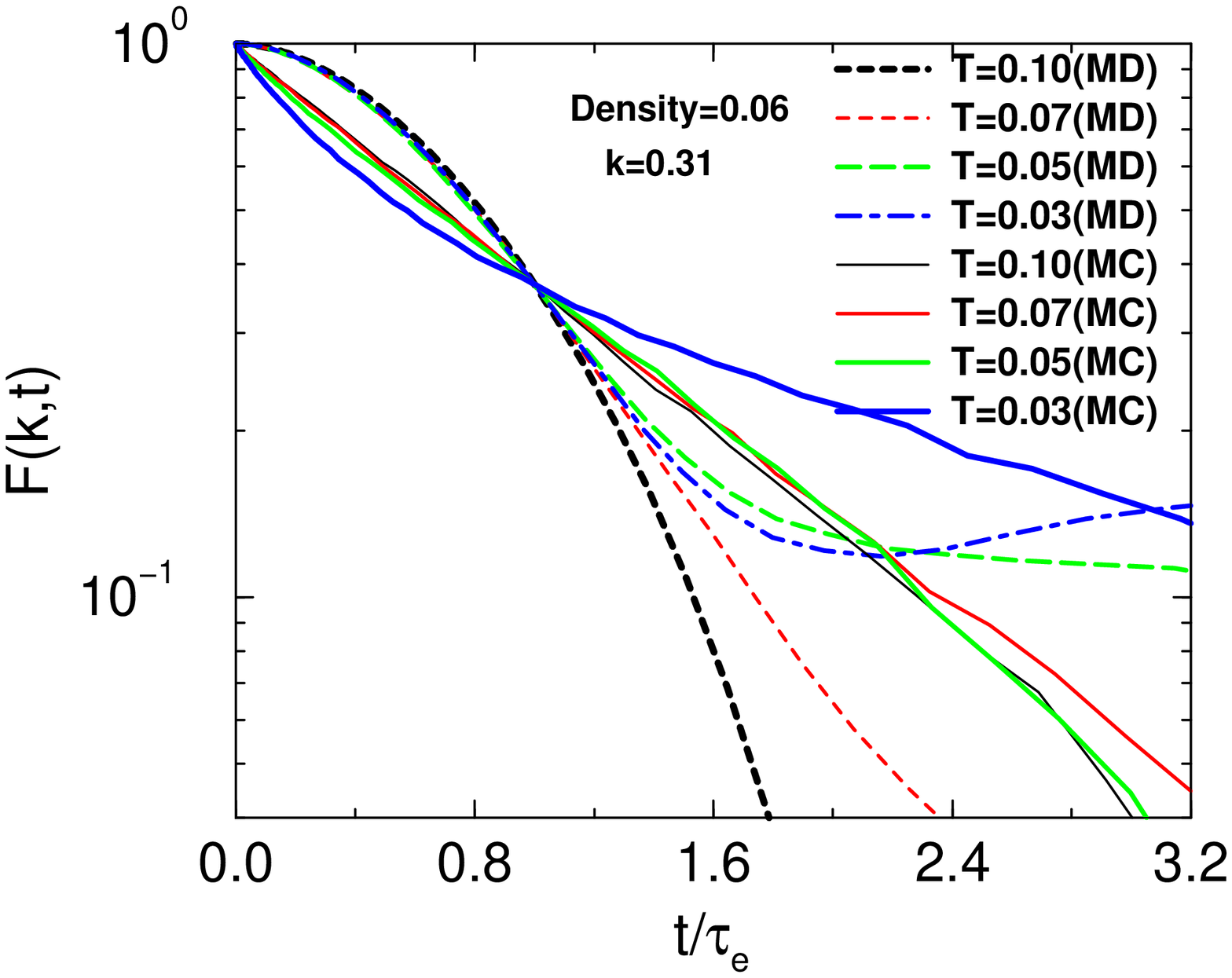}
\caption{The collective intermediate scattering function $F(k,t)$ {\it
vs.} time, for density = $0.06$, in log-linear scale as obtained from MD
(dashed lines) and MC (solid lines) for  $k=0.31$.}
\label{MC-Fkt-t-log-lin}
\end{figure}

In Fig.~\ref{MC-Fkt-t-log-lin} we show the coherent intermediate
scattering function $F(k,t)$ {\it vs.} time for different $T$'s 
and $k=0.31$ obtained from MC as well as MD simulations. In order to allow to
show the curves for different temperatures on the same graph, we plot the
correlators as a function of $t/t_{\rm e}$, where $t_{\rm e}$ is the time
at which the $F(k,t)$ reaches a value of $1/e$. From the graph we see
that the relaxation behavior of $F(k,t)$ from the MC and MD simulations
are quite different: Independent of $k$ \cite{EPAPS}, the $F(k,t)$ from MC shows an
exponential relaxation at the highest temperature shown ($T = 0.10$),
and becomes progressively more stretched as temperature decreases.
On the other hand, MD simulations exhibit compressed exponential for
all $T$, although a slower decay is apparent at lower $T$ due to the
presence of acoustic modes.  Thus we conclude that the MC dynamics
suppresses the intermediate time compressed exponential behavior and
allows one to see the stretched exponential decay at long times (see
also Ref.~\cite{shibuPRL}).

\begin{figure}[h]
\includegraphics[scale=0.50,angle=0]{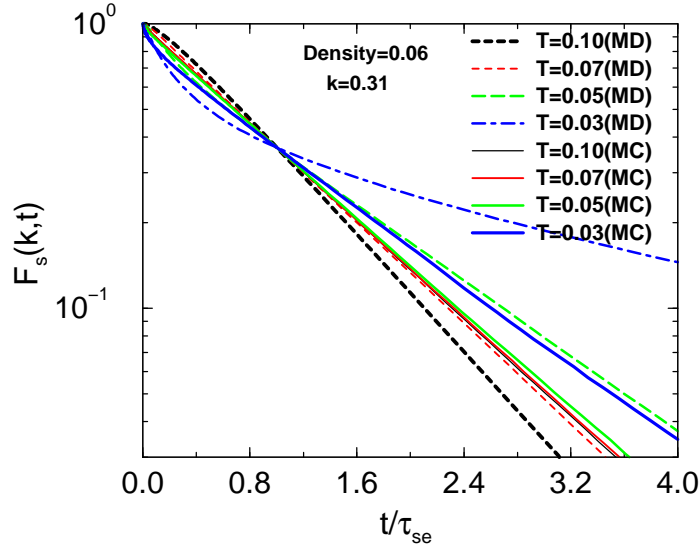}
\caption{The self intermediate scattering function $F_s(k,t)$ {\it vs.}
time, for density = $0.06$, in log-linear scale as obtained from MD
(dashed lines) and MC (solid lines) for  $k=0.31$.}
\label{MC-Fskt-t-log-lin}
\end{figure}

Last but not least we show in Fig. \ref{MC-Fskt-t-log-lin} the incoherent
intermediate scattering function $F_s(k,t)$ {\it vs.} $t/t_{\rm se}$
for different $T$ and $k$ obtained from MC as well as MD simulations.
The $F_s(k,t)$ curves from MC are nearly exponential at the lower $k$ values
and higher $T$, showing stretching at the lowest $T$ for all $k$ values,
and for all $T$ at $k=1.22$ \cite{EPAPS}. The degree of stretching increases with
decrease of $T$ and an increase of $k$. In contrast to this, $F_s(k,t)$
from MD shows a compressed exponential behavior for all $T$ at $k =
1.22$ \cite{EPAPS}, but at smaller $k$ the behavior becomes stretched at the 
two lowest temperatures.

\begin{figure}[h]
\includegraphics[scale=0.50,angle=0]{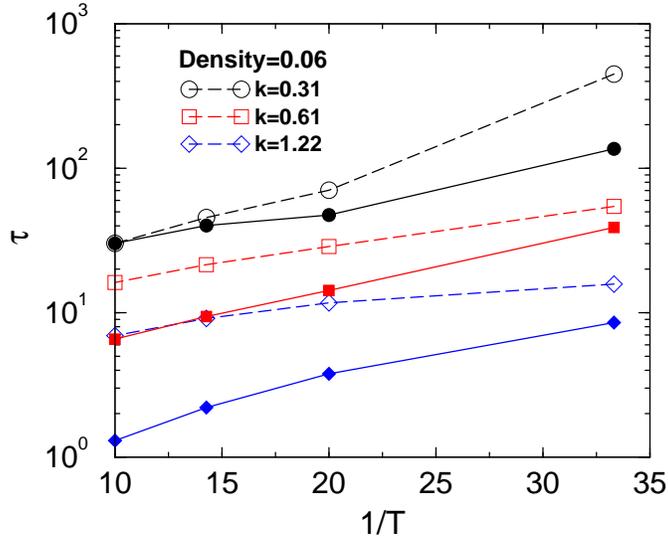}
\caption{The $T-$dependence of the relaxation times obtained from the area
under the coherent intermediate scattering function in MC (filled
symbols) and MD (open symbols) simulations at density $\rho=0.06$ for
different wave-vectors. The MC data are scaled by a factor $2000$ 
chosen such that the MC relaxation time at $k=0.31$
coincides with the corresponding MD time at the highest temperature.}
\label{MD-MC-tau-inverseT}
\end{figure}

Figure~\ref{MD-MC-tau-inverseT} shows the $T-$dependence of $\tau$
obtained from the area under $F(k,t)$ from MD and MC simulations for
wave-vector $k=0.31$, $0.61$ and $1.22$ at density $0.06$. The data
in Fig. \ref{MD-MC-tau-inverseT} have been plotted by rescaling the MC
relaxation times to be the same at $k = 0.31$ for the highest $T$. We
see that at $T=0.1$ the $k-$dependence of $\tau$ is much stronger for
the MD than for the MC. Furthermore we recognize that the $T-$dependence
of $\tau$ is stronger in the case of the MD than for the MC. Thus, the
long time behavior as revealed by MC is substantially different from
the self and collective intermediate scattering functions obtained from
MD. This result also implies that from the point of view of simulations it
is more efficient to equilibrate the system using MC.

\section{Summary and conclusions}

In this paper we have investigated the static and dynamic properties
of a recently introduced model for gel-forming systems. The
gel-forming ability of this model is related to the fact that its
local three-body potential avoids the formation of dense local
structures and instead favors an open network structure. Due to this
feature, the liquid-gas phase separation is pushed to low densities
and temperatures, thus opening at intermediate densities a
temperature range in which the relaxation dynamics is very slow,
i.e. the system is a gel {\it in equilibrium}.  Although this
mechanism for the formation of the gel is similar to the approaches
proposed earlier~\cite{delgado,fsrev,Blaak}, the simplicity of the
chosen interaction potential makes the present model very
attractive for simulations. This advantage is enhanced even
further by the use of an efficient method, presented in the Appendix,
to evaluate the mentioned three-body potential. Making use of this
computational gain we have been able to characterize in detail the
properties of the system, {\it in equilibrium}, in a significant range
of temperature and density.

We find that with decreasing temperature the particles assemble in the
form of long one-dimensional chains that are connected to each other at
random points, i.e. the system does indeed form a very heterogeneous
network in which the length of the bridging chains follows an exponential
distribution. The static structure factor of this spanning open network
shows a strong increase at small wave-vectors for temperatures that are
close to, but not quite at, the percolation line. The reason for this
shift, which depends on density, is the fact that at large scales $S(k)$
is not only given by the open structure of the percolating network but
that there are also contributions from the form-factor of the chains.

The bond correlation function $\phi_B(t)$ shows at moderately low
temperatures and high densities a non-exponential time dependence, which
is related to the complex relaxation dynamics on the length scale of two
particles, in analogy to the relaxation dynamics of dense glass-forming
liquids. In contrast to this, $\phi_B(t)$ shows at very low temperatures
and densities an exponential $t-$dependence, indicating a very simple
bond breaking mechanism with an activation energy that is only a very
weak function of $T$.  We note that this result is independent of the
used microscopic dynamics (molecular dynamics or Monte Carlo), which
shows that this dynamical property is closely related to the structure
of the system, as it has been found in other glass-forming systems
\cite{gleim,berthier,berthier_kob_07}.

In contrast to this, the time correlation functions $F(k,t)$ and $F_s(k,t)$
show a significant dependence on the microscopic dynamics. For $F(k,t)$
as obtained from the MD we find a compressed exponential for length scales
that are smaller or comparable to the one of the size of the particles,
and acoustic modes for larger scales. On the other hand, the MC dynamics
shows stretched exponentials for all length scales. The same trend is
seen for $F_s(k,t)$, except that there are (of course) no acoustic modes
in the MD. Thus we conclude that a dissipative dynamics, relevant for
the experimental systems, does {\it not} show the compressed exponential
relaxation observed in experiments {\it in which the sample was aging}. In
experiments, therefore, the out-of-equilibrium nature of the samples
must play an important role for the compressed exponential relaxation. Investigating the
details of how the out-of-equilibrium dynamics leads to the experimentally
observed compressed exponential relaxation remains an open problem that
should be investigated in the future.


\begin{acknowledgments}
We thank F. Sciortino, C. Pierleoni, J. Douglas, L. Cipelletti,
L. Berthier, M. Muthukumar, and A. Giacometti for fruitful discussions. We
acknowledge financial support from Indo-French Centre for the
Promotion of Advanced Research - IFCPAR, and CCMS, JNCASR for
computational facilities. Shibu Saw acknowledges CCMS, JNCASR for
support. We thank the Kavli Institute for Theoretical Physics,
U. C. Santa Barbara, for hospitality to Srikanth Sastry and W. K.,
providing an opportunity for this manuscript to be finalized. W.K. is
a senior member of the Institut universitaire de France. S. S. is
adjunct faculty at the International Centre for Theoretical Sciences,
TIFR.
\end{acknowledgments}

\section*{Appendix: Efficient method for force calculation of the SW potential}

Here we describe an efficient method for the calculation of energy and
forces for the Stillinger-Weber potential, which involves only double
sums to calculate three-body interaction energies and forces. We
follow the approach by Makhov and Lewis\cite{Makhov-Lewis} for energy
calculations (see also \cite{Frenkel-2loop,W-S1993}) but extend it to
the calculation of forces.

\subsection{Energy Calculation}

From Eqs.~(\ref{eq1})-(\ref{eq4}) it follows that the Stillinger-Weber
potential can be written as

\begin{equation}
u_{SW} = \sum_{i} \sum_{j> i} {v}_2 (r_{ij}) + \sum_{i} \sum_{j \ne i} \sum_{k \ne i,j} \frac{1}{2} 
\phi({\bf r}_{ij}, {\bf r}_{ik}) \quad
\label{eq13}
\end{equation}

\noindent
with distances expressed in units of $\sigma$ and $\phi$ denotes

\begin{eqnarray}
\phi({\bf r}_{ij}, {\bf r}_{ik}) &=& \epsilon \lambda \exp\Big(\frac{\gamma}{r_{ij}-a} + \frac{\gamma}{r_{ik}-a}\Big) (\cos \theta_{jik} +\alpha)^2 \nonumber \\
&& \times H(a-r_{ij}) H(a-r_{ik}) 
\label{eq14}
\end{eqnarray}

\noindent
We define $\tilde{\lambda} = \epsilon \lambda$, and
the quantities

\begin{eqnarray}
g_{ij} &=&
\left\{
\begin{array}{ll}
\exp \left[ \frac{\gamma}{r_{ij}-a}  \right] \quad & r_{ij} < a \\
0                                                  & r_{ij} \geq a
\end{array}
\right . ,  
\end{eqnarray}
\noindent

and

\begin{eqnarray}
U_{i}^{d} &=&  \tilde{\lambda} 
\sum_{j \ne i,k \ne i} \frac{1}{2} g_{ij} \mbox{ } g_{ik} (\cos \theta_{jik} + \alpha)^2 
\label{eq16}\\
&=& \tilde{\lambda}
\sum_{j \ne  i,k \ne i} \Big[ \frac{1}{2} g_{ij}  \mbox{ } g_{ik} \alpha^2 + 
\alpha g_{ij} \mbox{ }  g_{ik} \mbox{ } \cos \theta_{jik} \nonumber \\
&& + \frac{1}{2} g_{ij} \mbox{ } g_{ik} \mbox{ }\cos^2 \theta_{jik}     \Big] \quad .
\qquad \qquad \qquad \qquad \quad
\label{eq17}
\end{eqnarray}

With these definitions the three-body potential term in Eq.~(\ref{eq13}) can be written
as a sum over $U_i^d$,

\begin{equation}
\sum_{r_{ij},r_{ik}<a} \frac{1}{2} \phi({\bf r}_{ij}, {\bf r}_{ik}) =
\sum_i (U_i^d-{U_{ij}^c}') \quad ,
\label{eq18}
\end{equation}

\noindent
where the correction ${U_{ij}^c}'$ is due to the fact that $U_i^d$ contains
the terms $j=k$, see Eq.~(\ref{eq16}), whereas the sum on the left hand
side of Eq.~(\ref{eq18}) does not include them.

One easily finds that

\begin{equation}
{U_{ij}^c}'= \sum_{j\ne i} \frac{1}{2} \tilde{\lambda} g_{ij}^{2} \mbox{ } (1+\alpha)^2
\quad ,
\label{eq19}
\end{equation}

\noindent
i.e.~ this is a {\it two}-body term.

Defining thus 

\begin{equation}
v_{ij}^{eff}= v_2(r_{ij}) - U_{ij}^c \quad ,
\label{eq20}
\end{equation}

\noindent
where $U_{ij}^c = 2 {U_{ij}^c}'$,
the total potential energy becomes

\begin{equation}
u_{SW} =  \sum_{i} \sum_{j>i} v_{ij}^{eff}  + \sum_{i} U_{i}^{d}.
\label{eq21}
\end{equation}

The factor $1/2$ in Eq.~(\ref{eq19}) is to account for the double counting of pair distances,
which is not present in the first term of Eq.~(\ref{eq21}) and is, therefore, omitted.
In the following we will show that the term $U_i^d$ can be evaluated by
a {\it single} loop over the neighbors of particle $i$, thus allowing
us to evaluate the energy of the system without evaluating a triple sum.
For this we define the quantity $h_i$ as

\begin{equation}
h_i =  \sum_{j \ne i} g_{ij},  
\label{eq22}
\end{equation}

\noindent
and find 

\begin{equation}
h_{i}^{2} = \left| \sum_{j \ne i} g_{ij} \right|^2  
= \sum_{j,k \ne i} g_{ij} \mbox{ } g_{ik} \quad .
\label{eq23}
\end{equation}

\noindent
which is proportional to the first term on the right hand side of Eq.~(\ref{eq17}).

Similarly we define
 
\begin{equation}
{\bf s}_i =  \sum_{j \ne i} g_{ij} \hat {\bf r}_{ij}
\label{eq24}   
\end{equation}

\noindent 
where $\hat {\bf r}_{ij}= {\bf r}_{ij}/r_{ij}$ to obtain

\begin{eqnarray}
|{\bf s}_{i}|^{2} &=& \left| \sum_{j \ne i} g_{ij} \mbox{ }  \hat {\bf r}_{ij} \right|^2  
= \sum_{j,k \ne i} g_{ij} \mbox{ } g_{ik} (\hat {\bf r}_{ij} . \hat {\bf r}_{ik}) \\ 
&=& \sum_{j,k \ne i} g_{ij} \mbox{ } g_{ik} \mbox{ } \cos \theta_{jik} .
\label{eq25}
\end{eqnarray}

Finally, we define ${\bf T}_i$

\begin{equation}
{\bf T}_i = \sum_{j \ne i} g_{ij} (\hat {\bf r}_{ij} \otimes \hat {\bf r}_{ij} ) \quad ,
\label{eq26}
\end{equation}

\noindent
where $\otimes$ denotes the outer product, and obtain

\begin{eqnarray}
Tr[{{\bf T}_i}^2] &=&  \sum_{j,k \ne i} Tr[g_{ij} (\hat {\bf r}_{ij} \otimes \hat {\bf r}_{ij}) 
g_{ik} (\hat {\bf r}_{ik} \otimes \hat {\bf r}_{ik}) ]  \\
&=& \sum_{j,k \ne i} g_{ij} g_{ik}  (\hat {\bf r}_{ij}. \hat {\bf r}_{ik})^2 \nonumber \\ 
&=& \sum_{j,k \ne i} g_{ij} g_{ik}  \cos^2 \theta_{jik} .
\label{eq27}
\end{eqnarray}

\noindent 
Thus we find that $U_i^d$ is a sum of three terms, each of which
involves only a single sum over the neighbors of particle $i$:

\begin{equation}
U_i^d = \frac{\tilde{\lambda}}{2} \alpha^2 h_i^2 + \tilde{\lambda} \alpha |{\bf s}_i^2| + 
\frac{\tilde{\lambda}}{2} Tr[{\bf T}_i^2]
\label{eq28}
\end{equation}

This expression thus allows us to calculate efficiently the total potential
energy of the system using a double sum.

\subsection{Force Calculation}
The force acting on the $i^{th}$ particle is given  by
\begin{eqnarray}
{\bf F}_i &=& -\nabla_i u_{SW} \nonumber \\
&=& -\Big[ \sum_{j \ne i} \frac{\partial v_2(r_{ij})}{\partial r_{ij}} \hat {\bf r}_{ij} + \nabla_i U_i^d + \sum_{j \ne i} \nabla_i U_j^d  \nonumber \\ 
&& - \sum_{j \ne i} \frac{\partial U_{ij}^c}{\partial r_{ij}} \hat {\bf r}_{ij}  \Big].  
\end{eqnarray}
From Newton's third law of motion, we have 
\begin{eqnarray}
\nabla_i U_i^d &=& - \sum_{j \ne i}  \nabla_j  U_i^d. 
\end{eqnarray}

\noindent
Thus,
\begin{eqnarray}
{\bf F}_i&=& -\Big[ \sum_{j \ne i} \frac{\partial v_2(r_{ij})}{\partial r_{ij}} \hat {\bf r}_{ij} +  \sum_{j \ne i} (\nabla_i U_j^d - \nabla_j  U_i^d)  \nonumber \\ 
&& - \sum_{j \ne i} \frac{\partial U_{ij}^c}{\partial r_{ij}} \hat {\bf r}_{ij}  \Big], \nonumber \\
&&  \label{force}
\end{eqnarray}
where derivatives of trivial two-body terms are given as, 
\begin{eqnarray}
\frac{\partial v_2(r_{ij})}{\partial r_{ij}} &=& - \epsilon A \Big[ 4B r_{ij}^{-5} + \frac{Br_{ij}^{-4}-1}{(r_{ij}-a)^2} \nonumber \\
&&  \times \exp \left(  \frac{1}{r_{ij}-a}\right)  \Big]    
\end{eqnarray}
and
\begin{eqnarray}
\frac{\partial U_{ij}^c}{\partial r_{ij}}  &=& 2 \tilde{\lambda}  (1+\alpha)^2 g_{ij} \frac{\partial g_{ij}}{\partial r_{ij}}.
\end{eqnarray}


The non-trivial part of the calculation is the calculation of $\nabla_j U_i^d$:

\begin{eqnarray}
\nabla_j U_i^d &=&  \frac{\tilde{\lambda}}{2} \alpha^2  \nabla_j h_i^2 + \tilde{\lambda} \alpha \nabla_j |{\bf s}_i^2| + \frac{\tilde{\lambda}}{2}\nabla_j  Tr[{\bf T}_i^2 ].  \label{uid1}
\end{eqnarray}

Each of the terms are evaluated below. 

\begin{itemize} 

 
\item[1.]  Calculation of $\nabla_j h_i^2$:

\begin{eqnarray}
\nabla_j h_i^2 &=& 2 h_i \nabla_j h_i  \\
&=& 2 h_i \nabla_j \sum_{j \ne i} g_{ij}   \\
&=& 2 h_i  \frac{\partial g_{ij}}{\partial r_{ij}}  \hat {\bf r}_{ij}   \\
&=& 2 \frac{h_i}{ {r}_{ij}}  \frac{\partial g_{ij}}{\partial r_{ij}}  {\bf r}_{ij} \label{hiterm1} 
\end{eqnarray}


\item[2.] {\bf Calculation of $\nabla_j |{\bf s}_i^2|$}:

\noindent We have $\nabla_j |{\bf s}_i^2|$ = $\nabla_j ({\bf s}_i. {\bf s}_i) 
$. The gradient of the dot product of two vectors {\bf X} and {\bf Y} is given by,

\begin{eqnarray}
\nabla ({\bf X}. {\bf Y}) = ({\bf X}. \nabla) {\bf Y} + ({\bf Y}. \nabla) {\bf X} &+& {\bf X} \times (\nabla \times {\bf Y}) \nonumber \\ 
&+& {\bf Y} \times (\nabla \times {\bf X}). 
\label{gradxy}
\end{eqnarray}

\noindent Therefore, $\nabla_j ({\bf s}_i. {\bf s}_i) = 2 (\nabla {\bf s}_i) \mbox{}  {\bf s}_i$. The gradient of
vector ${\bf s}_i$ is a tensor of rank two i.e. a matrix and is given as,

\begin{eqnarray}
\nabla_j {\bf s}_i  &=& \nabla_j (g_{ij} \hat {\bf r}_{ij})  \\
&=& \frac{\partial g_{ij}}{\partial r_{ij}} (\hat {\bf r}_{ij} \otimes \hat {\bf r}_{ij}) + g_{ij} \nabla_j(\hat {\bf r}_{ij})  \\
&=& \frac{\partial g_{ij}}{\partial r_{ij}} (\hat {\bf r}_{ij} \otimes \hat {\bf r}_{ij}) + \frac{g_{ij}}{r_{ij}} \left[ {\bf I} - \hat {\bf r}_{ij} \otimes \hat {\bf r}_{ij}     \right], 
\end{eqnarray}

\noindent where we have used the fact that,

\begin{eqnarray}
\nabla_j \hat {\bf r}_{ij} = \frac{1}{r_{ij}} ({\bf I} - \hat {\bf r}_{ij} \otimes \hat {\bf r}_{ij}).
\end{eqnarray}

\noindent 
Hence,
 
\begin{eqnarray}
\nabla_j {\bf s}_i &=& \frac{g_{ij}}{r_{ij}} {\bf I} + \left( \frac{\partial g_{ij}}{\partial r_{ij}} - \frac{g_{ij}}{r_{ij}}  \right) (  \hat {\bf r}_{ij} \otimes \hat {\bf r}_{ij}  ).
\end{eqnarray}

\noindent  
Now, the gradient of $|{\bf s}_i|^2$ is given by,
 
\begin{eqnarray}
\nabla_j |{\bf s}_i|^2 &=& 2  (\nabla_j {\bf s}_i) \mbox{ } {\bf s}_i   \\
&=& 2 \left(\frac{\partial g_{ij}}{\partial r_{ij}} - \frac{g_{ij}}{r_{ij}} \right) (  \hat {\bf r}_{ij} \otimes \hat {\bf r}_{ij}  ) \mbox{ } {\bf s}_i \nonumber \\ 
&+& 2 \frac{g_{ij}}{r_{ij}}  {\bf s}_i 
\end{eqnarray}

\noindent where we  have used the fact that, 

\begin{eqnarray}
{\bf I} \mbox{ } {\bf s}_i &=& {\bf s}_i  \\ 
\mbox{and,   }  (\hat {\bf r}_{ij} \otimes \hat {\bf r}_{ij}) {\bf s}_i &=& (\hat {\bf r}_{ij} .{\bf s}_i ) \hat {\bf r}_{ij}  \\
&=& \frac{1}{r_{ij}^2} ( {\bf r}_{ij} .{\bf s}_i ) {\bf r}_{ij}. 
\end{eqnarray}

\noindent
Finally, the gradient of $|{\bf s}_i|^2$ is written as,

\begin{eqnarray}
\nabla_j |{\bf s}_i|^2 &=&  2 \left(\frac{\partial g_{ij}}{\partial r_{ij}} - \frac{g_{ij}}{r_{ij}} \right) \left( \frac{ {\bf s}_i. {\bf r}_{ij}} {r_{ij}^2}  {\bf r}_{ij} \right)  + 2 \frac{g_{ij}}{r_{ij}}  {\bf s}_i. \label{siterm1}
\end{eqnarray}

\item[3.]  Calculation of  $\nabla_j  Tr[{\bf T}_i^2]$:

We use the fact that
\begin{eqnarray}
\frac{\partial x^n}{\partial x} = n \frac{\partial (xA^{n-1})}{\partial x} \Bigg|_{A=x}.
\end{eqnarray}

\noindent 
Therefore, for a constant matrix {\bf A},

\begin{eqnarray}
\nabla_j Tr[{\bf T}_i^2] &=& 2 \nabla_j Tr({\bf T}_i{\bf  A}) \Big|_{{\bf A}={\bf T}_i}  \\
&=& 2 \nabla_j Tr[g_{ij} (\hat {\bf r}_{ij} \otimes \hat {\bf r}_{ij} {\bf A}) ]\Big|_{{\bf A}={\bf T}_i}   
\end{eqnarray}

\noindent One easily finds that, 

\begin{eqnarray}
Tr[ g_{ij} (\hat {\bf r}_{ij} \otimes \hat {\bf r}_{ij} {\bf A}) ]  = \Big( \frac{g_{ij} {\bf r}_{ij}. {\bf A} {\bf r}_{ij} }{ r_{ij}^2}    \Big). 
\end{eqnarray}

\noindent Therefore, the gradient of trace of $T_i^2$ matrix is given by 

\begin{eqnarray}
\nabla_j Tr[{\bf T}_i^2] &=& 2 \nabla_j \left( \frac{g_{ij} {\bf r}_{ij} . {\bf A}{\bf r}_{ij}}{r_{ij}^2}     \right) \Big|_{{\bf A=T}_i}  \\
&=& 2 \frac{\partial g_{ij}}{\partial r_{ij}}  \left( \frac{{\bf r}_{ij}. {\bf T}_i{\bf r}_{ij}}{r_{ij}^2} \right) \hat {\bf r}_{ij} \nonumber \\
&& + 2 g_{ij} \nabla_j \left( \frac{{\bf r}_{ij}. {\bf A}{\bf r}_{ij}}{r_{ij}^2}  \right) \\
&=& 2 \frac{\partial g_{ij}}{\partial r_{ij}} \left( \frac{{\bf r}_{ij}. {\bf T}_i{\bf r}_{ij}}{r_{ij}^2} \right) \hat {\bf r}_{ij} \nonumber \\
&& + 2 g_{ij}  (-\frac{2}{r_{ij}^3})  ( {\bf r}_{ij}. {\bf A}{\bf r}_{ij}) \hat {\bf r}_{ij}\Big|_{{\bf A=T}_i} \nonumber \\
&& + 2\frac{g_{ij}}{r_{ij}^2} \nabla_j ({\bf r}_{ij}. {\bf A}{\bf r}_{ij})\Big|_{{\bf A=T}_i}.
\end{eqnarray}

\noindent Using Eq.~(\ref{gradxy}), it is straightforward to show that

\begin{eqnarray}
\nabla_j ({\bf r}_{ij}. {\bf A}{\bf r}_{ij}) &=& 2 {\bf A}{\bf r}_{ij} 
\end{eqnarray}

\noindent because curl of ${\bf r}_{ij}$ and ${\bf A}{\bf r}_{ij}$ vanish, i.e., 

\begin{eqnarray}
\nabla \times {\bf r}_{ij}&=&0 \\
\nabla \times {\bf A}{\bf r}_{ij} &=& 0 \mbox{ (since {\bf A} is a symmetric matrix)}.
\end{eqnarray}

\noindent Therefore, finally the gradient of trace of $T_i^2$ matrix becomes 

\begin{eqnarray}
\nabla_j Tr\Big[ {\bf T}_i^2   \Big] &=& 2 \frac{\partial g_{ij}}{ \partial r_{ij}} \left(  \frac{{\bf r}_{ij} . {\bf T}_i {\bf r}_{ij}}  { r_{ij}^2}   \right)  \hat {\bf r}_{ij}  \nonumber \\ 
&&   - 4 g_{ij} \left(  \frac{{\bf r}_{ij} . {\bf T}_i {\bf r}_{ij}}  { r_{ij}^3}   \right) \hat {\bf r}_{ij} \nonumber \\
&& + 2 \frac{g_{ij}}{r_{ij}^2} ( 2 {\bf T}_i {\bf r}_{ij}   )   \\
&=& \left( 2\frac{\partial g_{ij}}{\partial r_{ij}}  - 4 \frac{g_{ij}}{r_{ij}}   \right)  \left( \frac{{\bf r}_{ij}. {\bf T}_i {\bf r}_{ij}}{r_{ij}^3}   \right) {\bf r}_{ij}  \nonumber \\ 
&&  + 4 g_{ij} \frac{{\bf T}_i{\bf r}_{ij}}{r_{ij}^2}.  
\label{Titerm1}
\end{eqnarray}

\end{itemize} 

\noindent Substituting Eqs. \eqref{hiterm1}, \eqref{siterm1}, and \eqref{Titerm1} into Eq. \eqref{uid1}, we get 

\begin{eqnarray}
\nabla_j U_{i}^d &=& \frac{\tilde{\lambda}}{2}\alpha^2 \left\{ 2 \frac{h_i}{r_{ij}}  \frac{\partial g_{ij}}{\partial r_{ij}} {\bf r}_{ij}     \right\}   +
\tilde{\lambda} \alpha \Big\{ 2 \Big( \frac{\partial g_{ij}}{\partial r_{ij}} - \nonumber \\
&&  \frac{g_{ij}}{r_{ij}} \Big)  \left(\frac{{\bf s}_i . {\bf r}_{ij}} {r_{ij}^2} \right) {\bf r}_{ij} + \frac{2 g_{ij}}{r_{ij}} {\bf s}_{i}    \Big \} +  \nonumber \\
&& \frac{\tilde{\lambda}}{2} \Big\{ \left(2\frac{\partial g_{ij}}{\partial r_{ij}} -4\frac{g_{ij}}{r_{ij}}\right)  \left(\frac{{\bf r}_{ij}. {\bf T}_i{\bf r}_{ij}}{r_{ij}^3} \right)  {\bf r}_{ij} +   \nonumber \\
&& 4 g_{ij} \frac{{\bf T}_i {\bf r}_{ij}}{r_{ij}^2}   \Big\}   \\
&=& c_{ij} {\bf r}_{ij} + 2 \tilde{\lambda} \alpha \mbox{ } \frac{g_{ij}}{r_{ij}} \mbox{ } {\bf s}_i + 2 \tilde{\lambda} g_{ij} \mbox{  } \frac{{\bf T}_i {\bf r}_{ij}}{r_{ij}^2},
\end{eqnarray}

\noindent where,

\begin{eqnarray}
c_{ij} &=& \left(\tilde{\lambda} \alpha^2 \frac{\partial g_{ij}}{\partial r_{ij}}  \frac{h_i}{r_{ij}}    \right) + 
2\tilde{\lambda} \alpha \left( \frac{\partial g_{ij}}{\partial r_{ij}} -\frac{g_{ij}}{r_{ij}} \right) \left( \frac{{\bf r}_{ij}. {\bf s}_i}{r_{ij}^2}\right)  +  \nonumber \\
 &&\tilde{\lambda} \left( \frac{\partial g_{ij}}{\partial r_{ij}} - \frac{2 g_{ij}}{r_{ij}}   \right) \left( \frac{{\bf r}_{ij} . {\bf T}_i{\bf r}_{ij}} {r_{ij}^3}  \right).
\end{eqnarray}

\noindent The total force on a particle is obtained from Eq. \eqref{force} as: 

\begin{eqnarray}
-{\bf F}_i &=& \sum_{j  \ne  i} \frac{\partial v_2}{\partial r_{ij}} \hat {\bf r}_{ij} - \sum_{j \ne i} \frac{\partial U_{ij}^c}{\partial r_{ij}} \hat {\bf r}_{ij}  + \nonumber \\
&&  \sum_{j\ne i} \left( \nabla_i U_j^d -\nabla_j U_i^d  \right)   \\
&=& \sum_{j \ne i} \frac{1}{r_{ij}}  \frac{\partial v_2}{\partial r_{ij}}  {\bf r}_{ij} -
\sum_{j \ne i} 2 \tilde{\lambda} (1+\alpha)^2 \frac{g_{ij}}{r_{ij}} \frac{\partial g_{ij}}{\partial r_{ij}}  {\bf r}_{ij} + \nonumber \\
&&   \sum_{j \ne i} (c_{ij} +c_{ji}) {\bf r}_{ij} +  \sum_{j \ne i} 2 \tilde{\lambda} \alpha \frac{g_{ij}}{r_{ij}} ({\bf s}_i -{\bf s}_j)+ \nonumber \\
&&   \sum_{j \ne i } 2 \tilde{\lambda} g_{ij} \frac{({\bf T}_i + {\bf T}_j)}{r_{ij}^2} {\bf r}_{ij} 
\end{eqnarray}

\noindent where 

\begin{eqnarray}
c_{ij} +c_{ji} &=& \tilde{\lambda} \alpha^2  \mbox{ } \frac{\partial g_{ij}}{\partial r_{ij}}  \mbox{ }  \frac{h_i+h_j}{r_{ij}} +
2 \tilde{\lambda}  \alpha \left(  \frac{\partial g_{ij}}{\partial r_{ij}} -\frac{g_{ij}}{r_{ij}}   \right) \nonumber \\
&& \times \frac{{\bf r}_{ij}. ({\bf s}_i-{\bf s}_j)}{r_{ij}^2} + \tilde{\lambda} \left( \frac{\partial g_{ij}}{\partial r_{ij}} - \frac{2g_{ij}}{r_{ij}}  \right) \times \nonumber \\
&& \frac{{\bf r}_{ij}.({\bf T}_i +{\bf T}_j)}{r_{ij}^3} {\bf r}_{ij}. 
\end{eqnarray}

This expression thus allows us to calculate efficiently the force on a particle, 
requiring only double sums. 


\clearpage

\begin{center}
\Large{{\bf Supplementary Material} }
\end{center}

\setcounter{page}{1}
\setcounter{figure}{0}
\setcounter{section}{0}

\makeatletter \renewcommand{\thefigure}{S\@arabic\c@figure} \renewcommand{\thesection}{S\@Roman\c@section}  \makeatother
\pagestyle{plain}

\section{Choice of Interaction Parameters $\alpha$ and $\lambda$}

\begin{figure}[b]
\vspace{10mm}
\includegraphics[scale=0.50,angle=0]{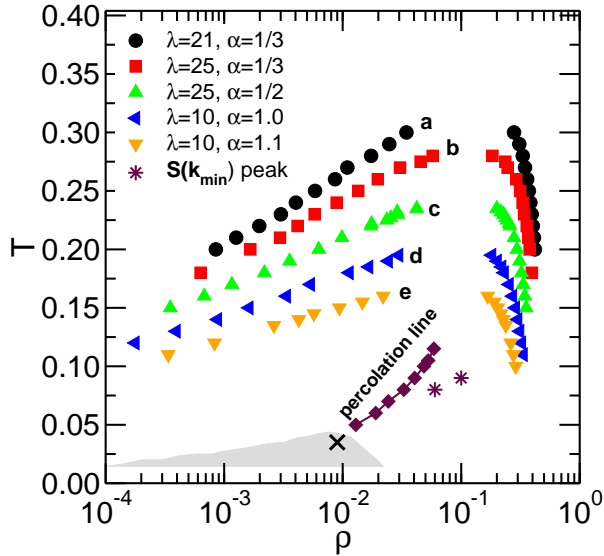}
\caption{The phase coexistence curves obtained from Gibbs Ensemble
Monte Carlo (GEMC) simulations in $T-\rho$ plane for (a) $\lambda$=21,
$\alpha$=1/3, (b) $\lambda$=25, $\alpha$=1/3, (c) $\lambda$=25,
$\alpha$=1/2, (d) $\lambda$=10, $\alpha$=1.0 and (e) $\lambda$=10,
$\alpha$=1.1. By increasing the value of $\lambda$ or $\alpha$, the phase
coexistence region shifts to low $T$ and $\rho$. The percolation line
is also shown.  The shaded region shows the estimated phase coexistence
region for the present model for $\lambda$=10, $\alpha$=1.49 based on MD
simulations (see text for details). The stars indicate the temperatures
(at constant density) at which the structure factor at the smallest
wave vector shows a maximum (see Fig. 2 of the main paper).
The cross indicates the density and temperature at which the $g(r)$
shown in Fig.  \ref{gofr-phasesep} is calculated.}
\label{phase-coex}
\end{figure}

We  describe the procedure by which the interaction parameters
$\alpha$ and $\lambda$ have been chosen. Figure \ref{phase-coex} shows
the LG phase-coexistence curves, obtained from Gibbs Ensemble Monte 
Carlo (GEMC) simulations\cite{Panagiotopoulos-gemc}, for various 
combinations of $\lambda$ and $\alpha$. 
The GEMC simulations have been performed with $2000$ particles, with maximum step size of $0.085$.
The values $\lambda=21$
and $\alpha=1/3$ represent silicon and the corresponding phase-coexistence
curve is the same  obtained by Honda {\it et al.}\cite{Honda}.
When $\lambda$ is increased to $25$ at constant $\alpha$
($=1/3$), we find that the LG phase-coexistence curve moves to lower
temperatures. Further, if $\alpha$ is increased to $1/2$ at fixed
$\lambda~ ($=25$)$, the LG phase-coexistence curve shifts to even lower
$T$'s.  If we keep $\lambda$ fixed at $10$ and increase $\alpha$, the LG
phase coexistence curve also moves to low $T$ and $\rho$ values. Thus
increasing either $\lambda$ or $\alpha$ results in diminishing the LG
coexistence region.

In addition to the effect on the LG coexistence, in order to ensure
that the state of the system at intermediate densities is stable with
respect to ordered structures (in our case, stackings of graphene-like
sheets), we choose (among other possible choices) values $\lambda = 10$
and $\alpha = 1.49$ (further details may be found in \cite{shibu-thesis-suppl}).

\begin{figure}[t]
\includegraphics[scale=0.50,angle=0]{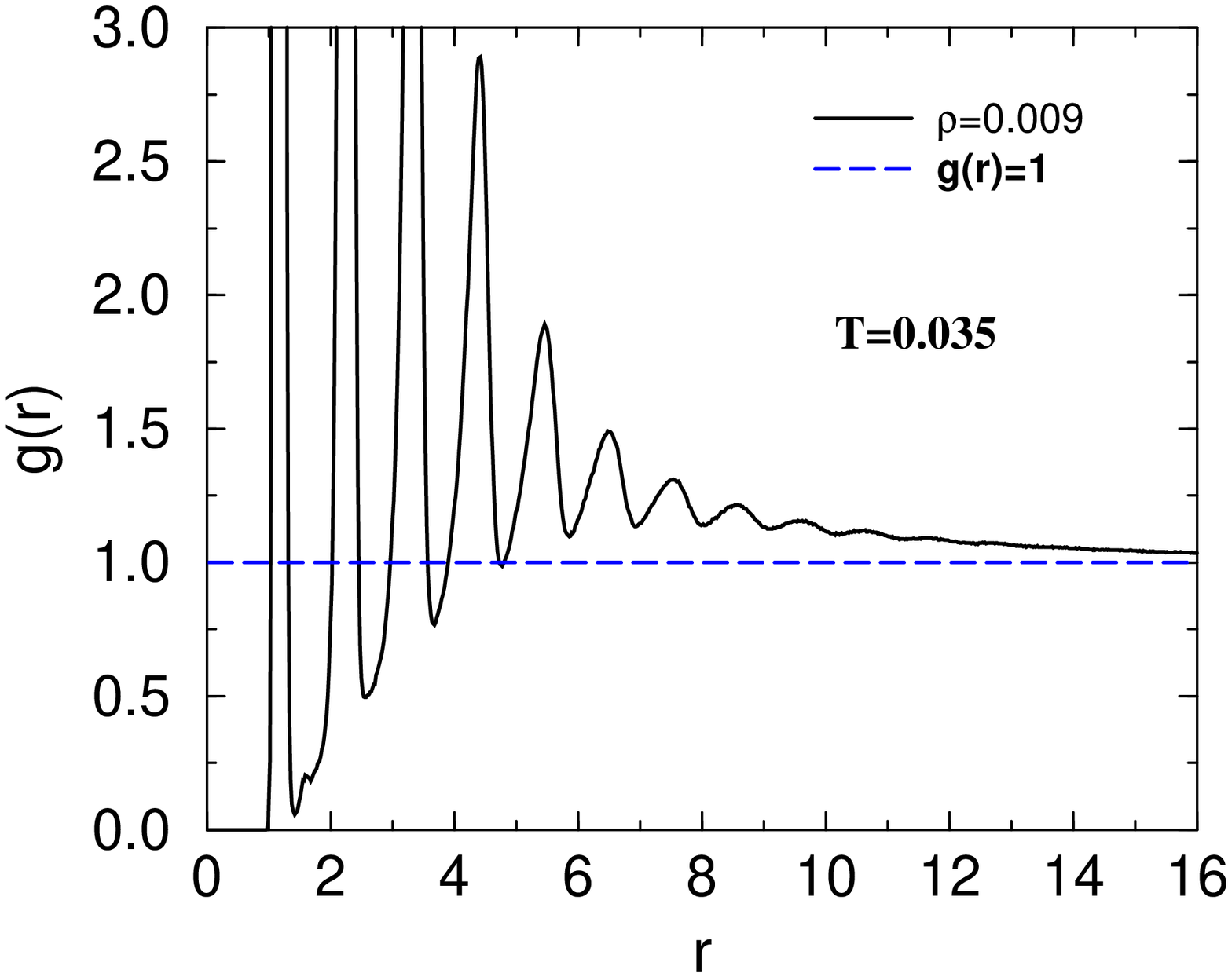}
\caption{The radial distribution function, $g(r)$, at density $\rho=0.009$
for $T = 0.035$ (indicated by a cross in Fig.~\ref{phase-coex}). The
$g(r)$ approaches its large-$r$ limit from above, which indicates
phase separation.}
\label{gofr-phasesep}
\end{figure}

The LG phase coexistence curve for the values $\lambda = 10$
and $\alpha = 1.49$ can not be determined
using GEMC simulations, because at low $T$ bonds become too strong
to swap particles between two sub volumes, which is required for GEMC
simulations. However, from Fig.~\ref{phase-coex} it is apparent that for
$\lambda=10$ and $\alpha=1.49$, the LG phase-coexistence curve will be
further suppressed compared to $\lambda=10$ and $\alpha=1.10$. A possible
way to deduce the location of the LG coexistence region is to look for
phase separation in a constant volume simulation. The shaded area in
Fig. \ref{phase-coex} shows the approximate region of LG phase coexistence
for the present model from such MD simulations. However, the densities of
both phases are very low, and therefore phase separation is not very easy
to determine from the MD snapshots. 
Nevertheless, phase separation can be deduced by studying the
radial distribution function, $g(r)$, which, if the system phase separates, 
approaches to $1.0$ from above
instead of oscillating around $1.0$ at large distances, the behavior found in 
homogeneous systems. This behavior
is shown in Fig.~\ref{gofr-phasesep}.

For completeness, we show in Fig.~\ref{phase-coex} also the percolation
line.  For a given temperature, the percolation density is determined
as the density at which the percolation probability, estimated from
considering several independent configurations, is $0.5$ (i.e.~$50\%$
of the configurations have a spanning cluster of bonded particles).
At the density $\rho = 0.06$, which we study in detail, the percolation
transition occurs at $T = 0.115$.

\section{Static Properties}

\begin{figure}[t]
\includegraphics[scale=0.60,angle=0]{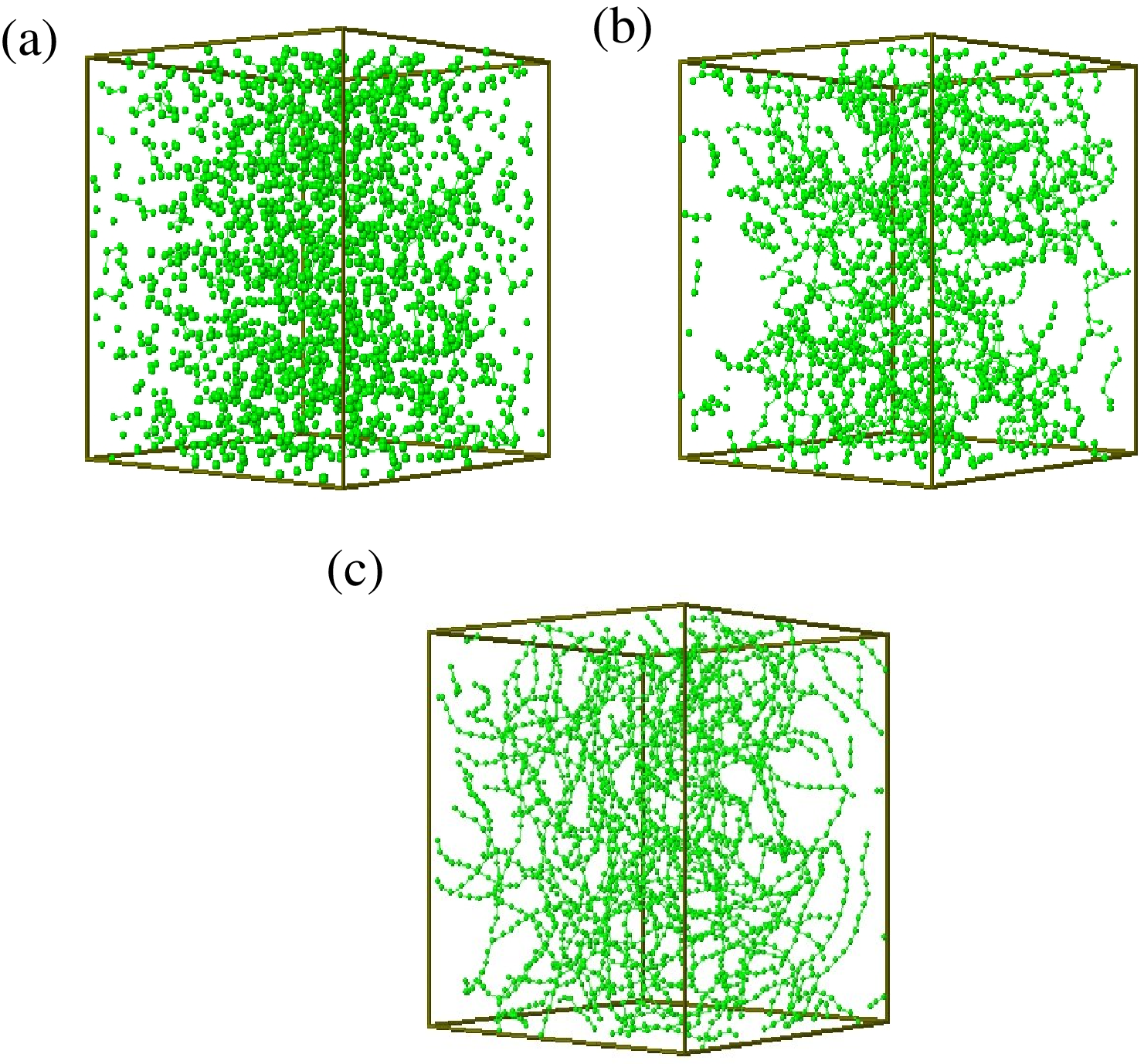}
\caption{Snapshots of the system from  MD simulations at density $0.06$
for: (a) $T=5.0$, (b) $T=0.08$, and (c) $T=0.028$.}
\label{snapshots}
\end{figure}

In order to get a first idea of the structure of the fluid, we show
in Fig.~\ref{snapshots} snapshots from the MD simulations at density
$0.06$ and different temperatures. At a very high temperature, $T =
5.0$ (Fig.~\ref{snapshots}a) the system is mainly composed of small
sized clusters. Upon lowering the temperature, particles form an
increasing number of interconnections, forming interlinked linear
chains that percolate if $T$ is below 0.115. Figure~\ref{snapshots}b
shows a snapshot at $T = 0.08$ manifesting large heterogeneity, and the
presence of significant voids, due to the proximity to the percolation
point. At still lower temperatures, particles form increasingly longer,
and stiffer, chains, and the system displays a more homogeneous morphology
(see Fig.~\ref{snapshots}c for $T = 0.028$).  Therefore, at the lowest
temperatures, the system consists mainly of linear chains of particles,
with a small number of three-fold coordinated (``anchor'') particles
connecting the linear chains.


\section{Wave-vector and Temperature Dependence of Dynamics}

\begin{figure}[h]
\includegraphics[scale=0.50,angle=0]{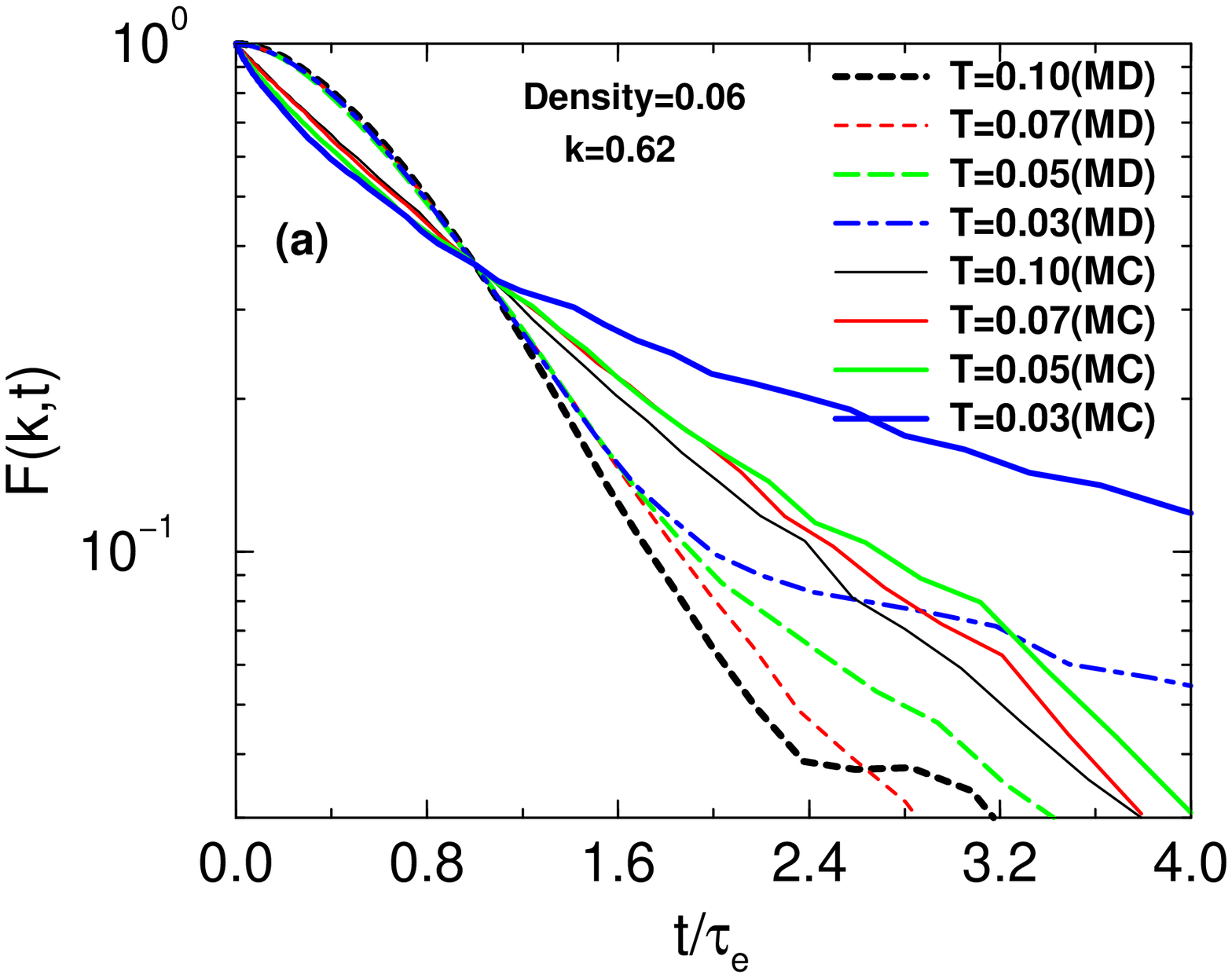}
\includegraphics[scale=0.50,angle=0]{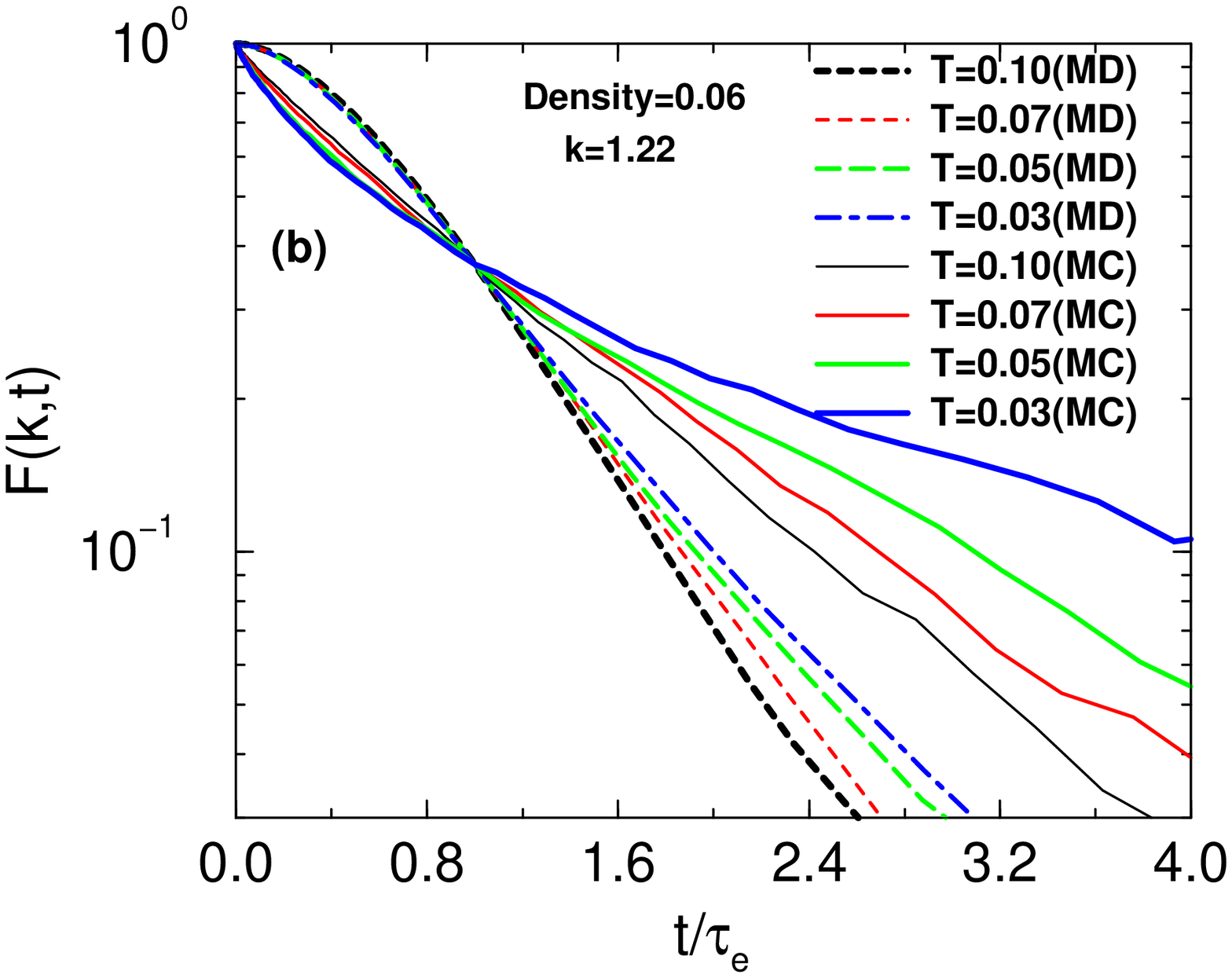}
\caption{The collective intermediate scattering function $F(k,t)$ {\it
vs.} time, for density = $0.06$, in log-linear scale as obtained from MD
(dashed lines) and MC (solid lines) for  (a) $k=0.62$ and
(b) $k=1.22$.}
\label{MC-Fkt-t-log-lin-suppl}
\end{figure}

In Fig.~\ref{MC-Fkt-t-log-lin-suppl} we show the coherent intermediate
scattering function $F(k,t)$ {\it vs.} time for different $T$ and $k$
values obtained from MC as well as MD simulations. In order to allow to
show the curves for different temperatures on the same graph, we plot the
correlators as a function of $t/t_{\rm e}$, where $t_{\rm e}$ is the time
at which the $F(k,t)$ reaches a value of $1/e$. From the graph we see
that the relaxation behavior of $F(k,t)$ from the MC and MD simulations
are quite different: Independent of $k$, the $F(k,t)$ from MC shows an
exponential relaxation at the highest temperature shown ($T = 0.10$),
and becomes progressively more stretched as temperature decreases.
On the other hand, MD simulations exhibit compressed exponential for
all $T$, although a slower decay is apparent at lower $T$ due to the
presence of acoustic modes.  Thus we conclude that the MC dynamics
suppresses the intermediate time compressed exponential behavior and
allows one to see the stretched exponential decay at long times (see
also Ref.~\cite{shibuPRL-suppl}).

\begin{figure}[h]
\includegraphics[scale=0.50,angle=0]{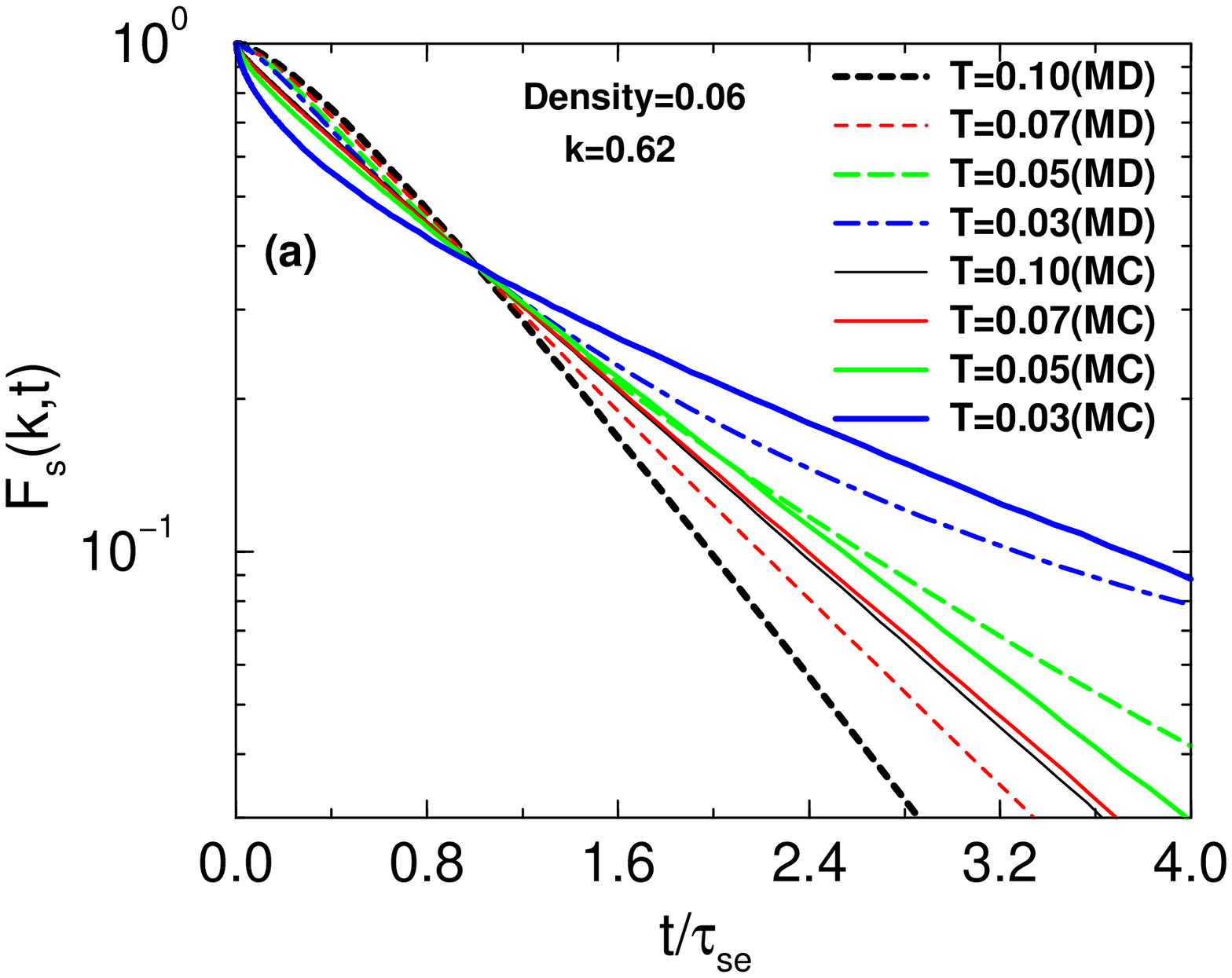}
\includegraphics[scale=0.50,angle=0]{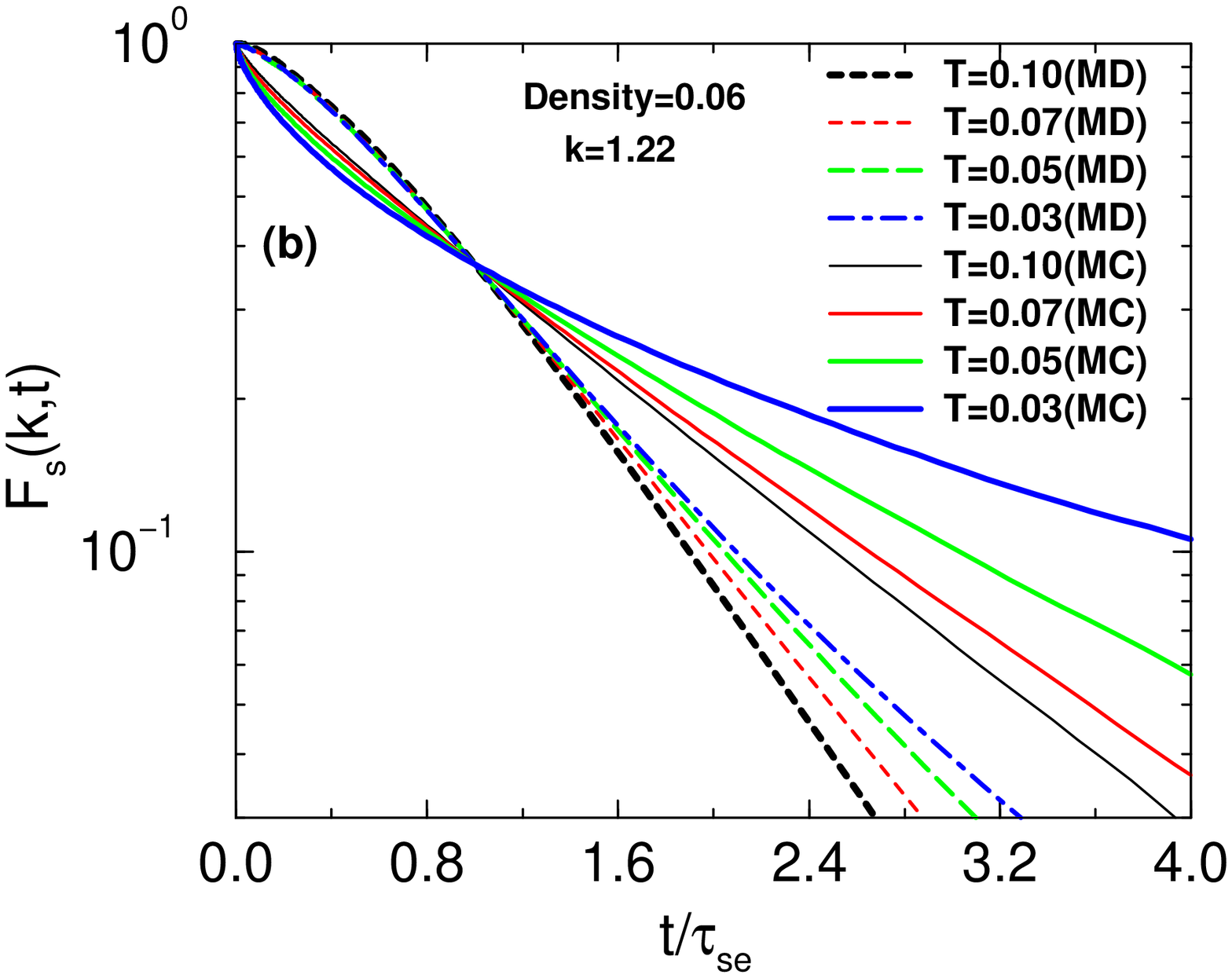}
\caption{The self intermediate scattering function $F_s(k,t)$ {\it vs.}
time, for density = $0.06$, in log-linear scale as obtained from MD
(dashed lines) and MC (solid lines) for  (a) $k=0.62$ and
(b) $k=1.22$.}
\label{MC-Fskt-t-log-lin-suppl}
\end{figure}

Last but not least we show in Fig. \ref{MC-Fskt-t-log-lin-suppl} the incoherent
intermediate scattering function $F_s(k,t)$ {\it vs.} $t/t_{\rm se}$
for different $T$ and $k$ obtained from MC as well as MD simulations.
The $F_s(k,t)$ curves from MC are nearly exponential at the lower $k$ values
and higher $T$, showing stretching at the lowest $T$ for all $k$ values,
and for all $T$ at $k=1.22$. The degree of stretching increases with
decrease of $T$ and an increase of $k$. In contrast to this, $F_s(k,t)$
from MD shows a compressed exponential behavior for all $T$ at $k =
1.22$, but at smaller $k$ the behavior becomes stretched at the 
two lowest temperatures (see main paper).

\section{Cluster Size Distribution}

\begin{figure}[h]
\includegraphics[scale=0.50,angle=0]{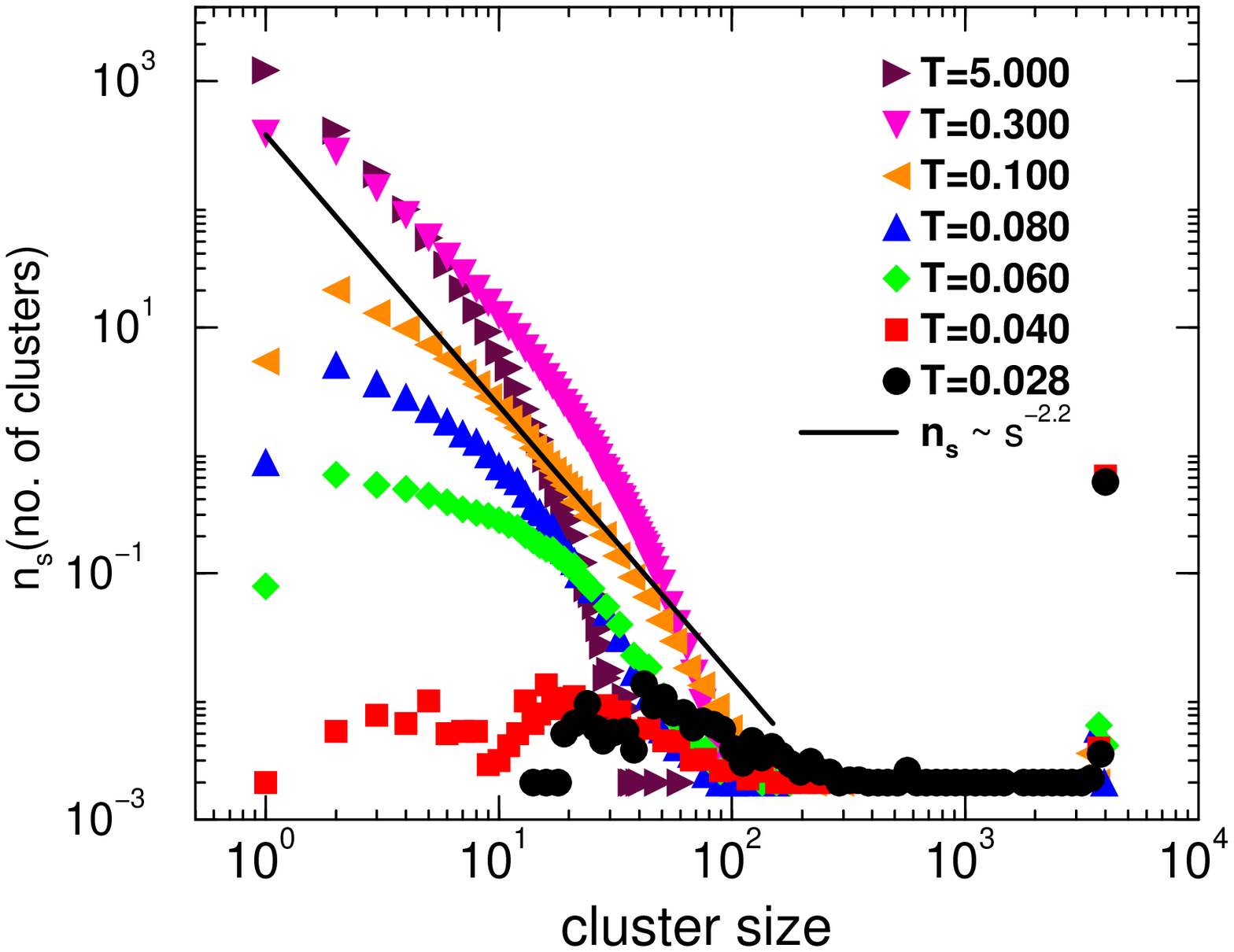}
\caption{The cluster size distribution $n_s$ for different temperature
  at density $0.06$. For very high temperature $T = 5.0$ the $n_s$
  shows an exponential form. At lower temperatures, including in the
  vicinity of the percolation transition, no distinguishable power-law
  regime is found, even though the cluster size distribution decays slower than exponentially.}
\label{ns-s}
\end{figure}

Fig. \ref{ns-s} exhibits the cluster size distribution $n_s$ for
different temperatures at density $0.06$. For very high temperatures
$e.g.$, $T = 5.0$, the $n_s$ shows an exponential decay, consistent
with the possibility of a a random aggregation process. When the
temperature is lowered, the distribution develops a slower than
exponential decay. However, even for temperatures near the percolation
transition, we do not find a distinguishable power-law regime, in
contrast to the findings for other gel forming systems for which a
power-law dependence corresponding to random percolation has been
observed \cite{KobEuro-suppl,Sciortino-ns-suppl,Coniglio-ns-suppl}.

\clearpage


\begin{thebibliography}{10}
\bibitem{footnote1} In order to perform constrained MD simulations,
we add a term of Gaussian form to the two-body potential that prevents
the breaking of existing bonds or the formation of new ones.  We use
the values 30, 1.40 and 0.0001 as the prefactor, mean and variance
of the imposed Gaussian potential. The Gaussian potential is shifted
appropriately to nullify the discontinuity in the force arising due to
truncation of the barrier potential.
\end{thebibliography}
\end{document}